\documentclass[a4paper,11pt]{article}
\usepackage{jcappub} 
\usepackage{lineno}
\usepackage{adjustbox}
\usepackage{arydshln}
\usepackage[final]{changes}

\arxivnumber{2410.15741} 
\title{\boldmath Detectability of Supernova Remnants with the Southern Wide-field Gamma-ray Observatory}

\author[a]{N. Scharrer,}
\author[a,b]{S.T. Spencer,}
\author[a]{V. Joshi,}
\author[a]{A.M.W. Mitchell}
\affiliation[a]{Erlangen Centre for Astroparticle Physics, Friedrich-Alexander-Universit\"at Erlangen-N\"urnberg,\\
Nikolaus-Fiebiger-Str. 2, 91058 Erlangen, Germany}
\affiliation[b]{Department of Physics, Clarendon Laboratory, Parks Road, Oxford, OX1 3PU, United Kingdom}

\emailAdd{nick.scharrer@fau.de}

\abstract{Supernova remnants (SNRs) are likely sources of hadronic particle acceleration within our galaxy, contributing to the galactic \replaced{c}{C}osmic \replaced{r}{R}ay flux. Next-generation instruments, such as the Southern Wide-field Gamma-ray Observatory (SWGO), will be of crucial importance in identifying new candidate SNRs. SWGO will observe two-thirds of the gamma-ray sky, covering the energy range between a few hundreds of GeV and a PeV. In this work, we apply a model of SNR evolution to a catalogue of SNRs in order to predict their gamma-ray spectra,  explore the SNR emission phase space, and quantify detection prospects for SWGO. Finally, we validate our model for sources observed with current-generation instruments, fitting it using a Monte-Carlo Markov Chain technique to the observed gamma-ray emission from four SNRs. We anticipate that at least 6, and potentially as many as 11 SNRs will be detected by SWGO within 1 year.}

\begin{document}
\maketitle
\flushbottom

\section{Introduction}
The Southern Wide-field Gamma-ray Observatory (SWGO) is proposed to be the next generation ground-based water-Cherenkov detector array to be built in South America. SWGO will be at an altitude of 4770\,m above sea level, making it an ideal ground-based gamma-ray sky survey observatory in an energy range from a few hundred GeV to a PeV. 
\added{Astrophysical g}{G}amma-rays\added{ in this energy range} can have either a leptonic or hadronic origin. Under the leptonic scenario, inverse Compton scattering of energetic electrons boost photons from background radiation fields into the gamma-ray regime. Under the hadronic scenario, proton-proton interactions generate charged and neutral pions, that rapidly decay producing muons, neutrinos and gamma-ray photons (the latter two both being neutral messengers).  Several known source classes can host the environment to produce gamma-rays of these energies \citep{Hinton_2009}. One of these source classes, Supernova-Remnants (SNRs), are likely to contribute to the acceleration of hadronic Cosmic Rays, although the maximum energy to which this is possible is still a matter of debate \citep{2023ApJ...958....3D,christofari}. \replaced{The gamma-ray emission from SNRs observable above 10\,TeV is likely to be hadronic rather than leptonic in nature \citep{aharoniangamma}.}{SNRs are also expected to produce hadronic gamma-ray emission.}

SWGO will survey the southern gamma-ray sky in a declination range of approximately $-70^{\circ}$ to $+20^{\circ}$ deg, encompassing most of the galactic plane. This will allow for observations of known southern sky SNRs at higher energies, and for detecting previously unknown SNRs. In this work, we aim to determine which of the catalogued SNRs will be detectable by SWGO. To achieve this, we use an SNR evolution and gamma-ray emission model to predict the gamma-ray spectrum emitted by different SNRs. We compare the predicted SNR spectra to the SWGO target sensitivity range \citep{2023arXiv230904577C}, and then validate our model with well-known SNRs to verify its robustness. This work therefore provides useful input to the SWGO planning for the science cases pertaining to SNRs.

\section{Model description}
\label{sec:model_considerations}
\replaced{In this section we will introduce the model we use to predict the gamma-ray emission from SNRs.}{To predict the gamma-ray spectra produced by SNRs, we devise a model taking into account the following considerations.} Firstly, we consider the SNRs' evolution to determine their properties such as their size and \replaced{age}{lifetime}, which are in turn of crucial importance to determine the particle spectrum they produce. \deleted{Our model also takes into account the system's history, making it time-dependent.} \added{We will then describe how we use the GAMERA software package \cite{2015ICRC...34..917Hahn,2022ascl.soft03007HahnGAMERA} to estimate the resulting gamma-ray emission produced by these particles.}\deleted{The evolution of an SNR is divided into four phases distinguished by their expansion velocity \cite{SupernovaRemnantsWoltjer}.}

\subsection{SNR evolution}
\added{The evolution of an SNR is divided into four phases distinguished by their expansion velocity \cite{SupernovaRemnantsWoltjer}.} SNRs begin their life with a so-called \textbf{ejecta-dominated phase}. The rate of expansion depends on the type of predecessor supernova type, ranging from $5000~\rm{km/s}$ for core-collapse supernovae to  $10^4~\rm{km/s}$ for type Ia supernovae. \replaced{We}{The latter occur when a white dwarf exceeds the Chandrasekhar mass limit, however we} only consider core-collapse supernovae in this work. For these we adopt the \replaced{model}{work} of \citep{CARDILLO20151} to \replaced{characterise}{model} the SNR's early evolution.

During the ejecta dominated phase for a core collapse supernova, the SNR expands into the wind of the progenitor star. When the SNR has swept up as much mass from the interstellar medium (ISM) as the initially ejected mass $M_{\rm ej}$\deleted{,} a state of equilibrium is reached, beginning the \textbf{Sedov-Taylor or energy conservation phase}. For a SNR expanding into the wind of its progenitor red giant, the corresponding radius at which this occurs is
\begin{equation}
    R_0 = \frac{M_\mathrm{ej}V_\mathrm{w}}{\dot{M}}\approx1\left(\frac{M_\mathrm{ej}}{M_\odot}\right)\left(\frac{V_\mathrm{w}}{10\,\mathrm{km\,s^{-1}}}\right)\left(\frac{\dot{M}}{10^{-5}M_\odot\,\mathrm{yr^{-1}}}\right)\,\mathrm{pc}
\label{eqn:radius_sedov_time}
\end{equation}
\citep{CARDILLO20151} which is reached at the so-called Sedov time
\begin{equation}
    t_{\mathrm{sed}} = \left[\left(\frac{B}{A}\right)^{1/(k-m)}\right]^\frac{k-m}{k-3},
\label{eqn:time_sedov_phase}
\end{equation}
where $k=9$ and $m=2$ for type II supernovae \citep{CARDILLO20151}.
The normalisation factor $A$ is given by
\begin{equation}
    A=\frac{1}{4\pi k}\frac{\left[3(k-3)M_{\mathrm{ej}}\right]^{5/2}}{\left[10(k-5)E_{\mathrm{ej}}\right]^{3/2}} \left[\frac{10(k-5)E_{\mathrm{ej}}}{3(k-3)M_{\mathrm{ej}}}\right]^{k/2},
\end{equation}
where $A$ arises from assuming self-similarity for the time evolution of the shock \citep{1999_Truelove_McKee}, and $B=\dot{M}/4\pi V_\mathrm{w}$, which arises from the condition that the density of the wind and ejecta are equal at the SNR forward shock \citep{CARDILLO20151}.
The development of the shock radius in both the ejecta dominated and Sedov-Taylor phases is then given by 
\begin{equation}
    R_\mathrm{sh}(t)= R_0 \left( \left( \frac{t}{t_{\mathrm{sed}}} \right)^{a\lambda_\mathrm{ED}} + \left( \frac{t}{t_{\mathrm{sed}}} \right)^{a\lambda_\mathrm{ST}} \right)^{1/a},
\label{eq:radius_sedov_phase}
\end{equation}
\citep{CARDILLO20151} with $\lambda_\mathrm{ED}=(k-3)/(k-m)$, $\lambda_\mathrm{ST}=2/(5-m)$ and the smoothing parameter that models the transition between the two phases $a=-5$. The shock velocity during these phases is then given by
\begin{equation}
    v_{\mathrm{sh}}(t)=\frac{R_{0}}{t_{\mathrm{sed}}}\left(\frac{R}{R_{0}}\right)^{1-a}\left[\lambda_\mathrm{ED}\left(\frac{t}{t_{\mathrm{sed}}}\right)^{a \lambda_{\mathrm{ED}}-1}+\lambda_\mathrm{ST}\left(\frac{t}{t_{\mathrm{sed}}}\right)^{a \lambda_{\mathrm{ST}}-1}\right].
\end{equation}
Until now, the energy of the system has only been distributed in thermal and kinetic energy. However, this is no longer true towards the end of the Sedov-Taylor phase. Because the radiative losses start to become non-negligible, the next stage of the SNR evolution is defined as the \textbf{snow-plough, pressure-driven phase or radiative phase}. The corresponding timescale can be defined as
\begin{equation}
    t_\mathrm{rad} = 1.4 \cdot 10^{12} \left( \frac{E_{51}}{\rho_0}\right)^{\frac{1}{3}} \mathrm{s} 
                    \approx 44600 \left( \frac{E_{51}}{\rho_0} \right)^{\frac{1}{3}}\,\mathrm{yr}
\label{eqn:time_rad_phase}
\end{equation}
\citep{vink2020physics}. For typical values $t_\mathrm{rad} \approx 100\,\mathrm{kyr}$. Nevertheless, the momentum of the system is still conserved, and the radius dependency on time decreases to $R(t) \propto t^{\frac{1}{4}}$.

Lastly, in about 10 million years, the velocity of the ejecta shell drops down to the sound speed in the ISM\replaced{. This results}{, resulting} in a constant radius $\approx100\,\mathrm{pc}$ \replaced{which marks}{marking} the end of the SNR's evolution, known as the \textbf{merging phase}.

\subsection{Particle acceleration}
\begin{figure}[htp]
\centering
\includegraphics[width=0.65\linewidth]{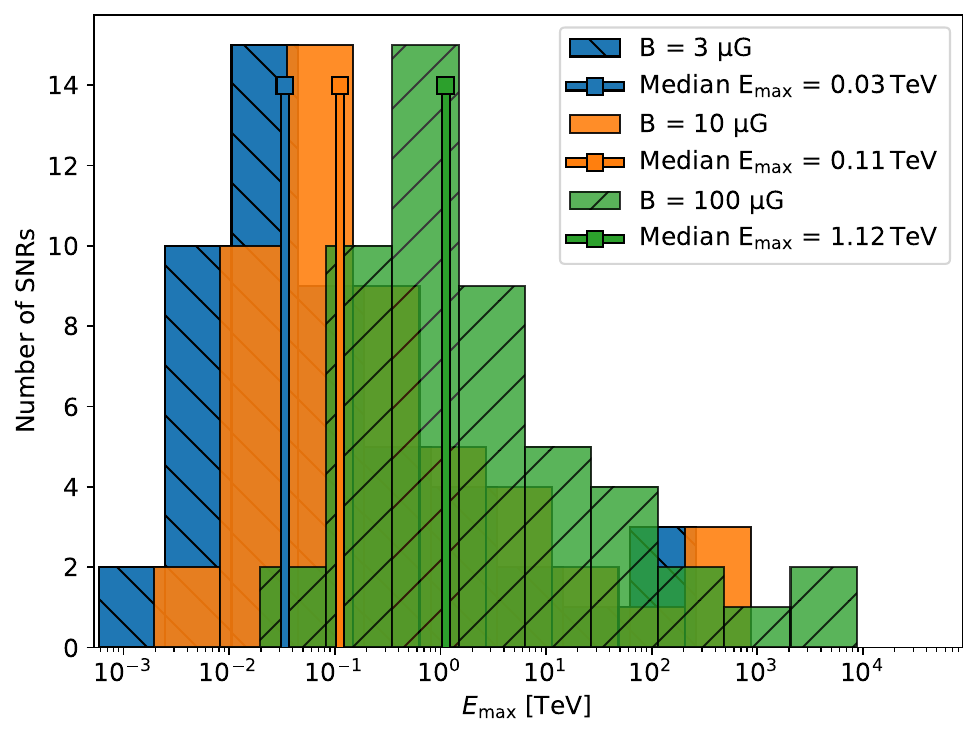}
\caption{Change of the distribution of maximum particle energies for SNRs with the magnetic field strength. } 
\label{fig:emax_calculation}
\end{figure}

In an SNR shock wave, particle acceleration is believed to be driven by first-order Fermi acceleration, also known as diffusive shock acceleration \citep{1978MNRASBell}. In this mechanism, particles are repeatedly scattered across the shock front due to turbulent magnetic fields, gradually increasing their energy. This process results in the observed power-law spectra. Their maximum energy is dependent on the magnetic field strength $B$ and the shock velocity $v_\mathrm{sh}$. This energy can be approximated following \citep{Blasi_2013} as
\begin{equation}\label{eq:emax}
    E_\mathrm{max}\approx 3\times 10^5\,\mathrm{GeV}\ B_{100}\left(\frac{t_\mathrm{sed}}{300\,\mathrm{yr}}\right)\left(\frac{v_\mathrm{sh}}{1000\,\mathrm{km\,s}^{-1}}\right)^2,
\end{equation}
where $B_\mathrm{100}$ is the magnetic field strength in units of 100\,$\mu$G. This equation assumes the maximum energy reached by diffusive shock acceleration, with magnetic field amplification upstream of the shock and Bohm-type diffusion, is highest at the end of the ejecta dominated phase (as this is the time at which the shock velocity is highest). It is obtained by comparing the acceleration timescale to the age of the SNR. Figure \ref{fig:emax_calculation} shows how the estimated $E_{\rm max}$ of SNRs changes with $B$. The median value increases from 0.15\,TeV to 5.16\,TeV with increasing magnetic field and the number of SNRs exceeding 1\,PeV is 0, 0 and 3 for magnetic field strengths of 3, 10 and 100\,$\mu$G respectively.

For the energy spectrum of the protons, we use a power-law with a cut-off \cite{Celli_2020}
\begin{equation}\label{eq:J_p}
    J_\mathrm{p}\equiv \frac{dN_\mathrm{p}}{dE_\mathrm{p}dV}=K_\mathrm{p}E_\mathrm{p}^{-\alpha_\mathrm{p}}\exp\left(-\left(\frac{E_\mathrm{p}}{E_{0,\mathrm{p}}}\right)^{\beta_\mathrm{p}}\right),
\end{equation}
setting the exponents to their typically used values of $\alpha_\mathrm{p}=2$ and $\beta_\mathrm{p}=1$ \citep{Celli_2020}. The normalisation constant $K_\mathrm{p}$ is derived from the energy density above 100\,GeV
\begin{equation}\label{w_p}
    w_\mathrm{p}=\int\displaylimits_{100\,\mathrm{GeV}}^{\infty}E_\mathrm{p}J_\mathrm{p}(E_\mathrm{p})dE_\mathrm{p}=1\,\mathrm{erg\,cm^{-3}}.
\end{equation}
The maximum energy $E_\mathrm{0,p}$ at the current age is determined by Equation \ref{eq:emax} at $t=t_\mathrm{sed}$\added{,} where the particles of highest energy $p_\mathrm{M}$ can escape the acceleration region
\begin{equation}\label{eq:p_M}
    p_{\mathrm{max,0}}(t)=p_\mathrm{M}\left(\frac{t}{t_\mathrm{sed}}\right)^{-\delta}\equiv E_{0,\mathrm{p}},
\end{equation}
\deleted{where }$\delta$ is a free parameter of the model that characterises the change in maximum energy over time at $t>t_{\rm sed}$ \cite{Celli_thesis}. 
This parameterisation of $p_{\mathrm{max,0}}(t)$ is a commonly used phenomenological approach to account for uncertainties associated with the magnetic turbulence generated by the accelerated particles themselves \citep{gabici09,Celli_thesis}. We assume that magnetic turbulence amplification takes place upstream of the shock, increasing the magnitude of the time-dependence $\delta$, for which we adopt $\delta=3$ as a baseline value. Using this value, it is possible to replicate the observed very-high-energy gamma-ray emission for certain SNRs \citep{Celli_thesis}. Assuming a power-law time dependence in this manner results in a power-law distribution of escaped particles that can come close to replicating the observed cosmic ray spectrum below the knee \citep{gabici09}. However, the value of $\delta$ is not well constrained \citep{Celli_thesis}, and we explore the phase space for $\delta$ in Section \ref{sec:modelbehaviour}.

\subsection{Gamma-ray production}
Proton-proton (p-p) interactions are believed to be the dominant process yielding gamma-ray emission from SNRs in the energy range relevant to SWGO \citep{aharoniangamma}. Energetic protons accelerated in the SNR shock environment have a spectrum (Equation \ref{eq:J_p}), and interact with target material in the surrounding medium. These interactions generate energetic pions as by-products, that rapidly decay into either electrons / positrons and neutrinos in the case of charged pions or into two gamma-ray photons in the case of neutral pions. To calculate the resulting Spectral Energy Distributions (SEDs) we use GAMERA, a library for particle and gamma-ray modeling of astrophysical sources \cite{2015ICRC...34..917Hahn,2022ascl.soft03007HahnGAMERA}. GAMERA incorporates gamma-ray emission arising from p-p interactions following the treatment of \cite{2014PhRvD..90l3014Kafexhiu}, which utilises parameterised approximations to the inelastic scattering cross-section $\sigma_{\rm inel} (E_p)$. The flux of gamma-ray photons with energy $E_\gamma$ produced at the source location due to p-p interactions 
can be written: 
\begin{equation}
\Phi(E_\gamma)=4\pi n_H \int \frac{d\sigma}{dE_\gamma}\left(E_p,E_\gamma\right)J(E_p)dE_p\,,
    \label{eq:gammaflux}
\end{equation}
where $n_H$ is the density of the target material with which the interactions occur, and $J_p(E_p)$ is the proton energy spectrum. 

\section{Model characteristics}

\label{sec:modelbehaviour}
In this section, we explore how free parameters of the model for SNR evolution and particle acceleration influence the resulting gamma-ray SED curves. We compare this to the target SWGO sensitivity band, for which we use public estimates from \citep{2023arXiv230904577C} throughout this work. The upper bound of this \deleted{shaded} sensitivity band corresponds to the baseline anticipated sensitivity of SWGO in a 1\,year exposure. \replaced{As development of detector technologies and array layout optimisation for SWGO is ongoing, a range of lower sensitivities are possible.}{Due to the ongoing development of detector technologies and array layout optimisation, the shaded sensitivity band indicates the phase space exploration for SWGO.} In particular,\deleted{ the} improvements at low energies ($\lesssim 1$\,TeV) and at high energies ($\gtrsim 50$\,TeV) can be achieved with compact and sparse large area array layouts respectively. Improvements in the core energy range could be achieved with improved PSF and background rejection efficiencies, through either detector technology or analysis algorithms. Table \ref{tab:parameters} summarises the default values for free parameters of our model, with the exception of the SNR age and distance, which are adapted for each individual SNR we consider. 

\begin{table}[!htbp]
    \centering
    \begin{tabular}{c|c|c|c}
    \hline
    Parameter & Assumed Value & Possible Range & Reference \\
    \hline
    $B$ & 100\,$\mu$G& 50 - 200 $\mu$G &\citep{annurev:/content/journals/10.1146/annurev.astro.46.060407.145237}\\
    $M_{\rm ej}$ & $3\,M_\odot$& $1-10\,M_\odot$ &\citep{Bamba_2022}\\
    $\rho_0$ & $1\,m_{\rm p}\,\rm cm^{-3}$& $0.1-100\,m_{\rm p}\,\rm cm^{-3}$ &\citep{annurev:/content/journals/10.1146/annurev.astro.46.060407.145237} \\
    $\delta$ & $3$& Poorly Constrained & \citep{Celli_thesis} \\
    $E_{\rm ej}$ & $10^{51}\,\mathrm{erg}$& $\sim0.1-10\times10^{51}\,\mathrm{erg}$&\citep{diesing,vink2020physics}\\
    $\alpha_{\rm p}$ & 2&Common Assumption&\citep{Celli_2020}\\
    $\beta$ & 1& Common Assumption&\citep{Celli_2020}\\
    \hline
    \end{tabular}
    \caption{Values adopted for free parameters of our SNR evolutionary model, unless otherwise specified. For the proton spectrum, we use an exponential cut-off power law model with index $\alpha_{\rm p}$ (Equation \ref{eq:J_p}). }
    \label{tab:parameters}
\end{table}

\begin{figure}[htp]
\centering
\includegraphics[width=0.7\linewidth]{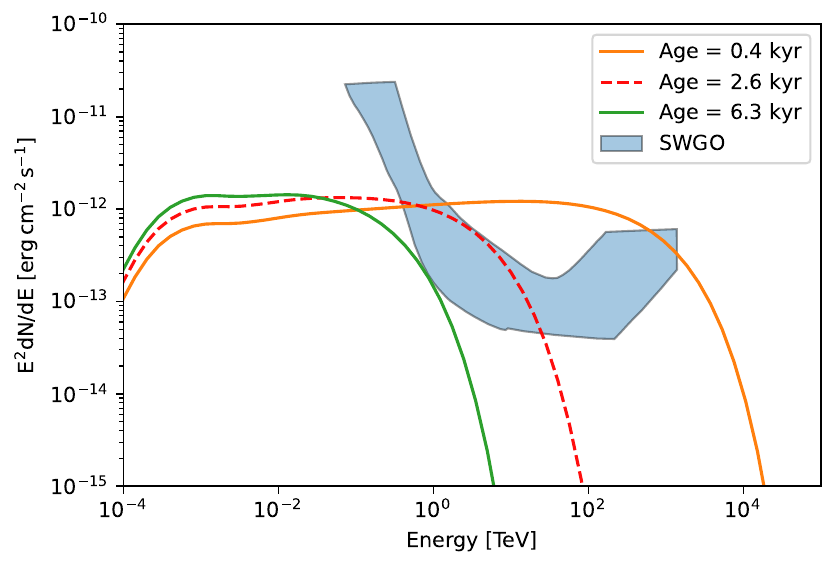}
\caption{The influence of the SNR age on the SED, the dashed red line is for the median age of the SNRs in the sample that are detectable, the orange line for the 5th and the green line for the 95th percentile of these SNRs \cite{SNRcat_article}. We assume here a distance of 3.7\,kpc, which is the mean value for the sources in the SNR sample that exceed the SWGO sensitivity. We take the SWGO sensitivity band from \citep{2023arXiv230904577C} for comparison.}
\label{fig:model_vary_age}
\end{figure}

Firstly, we note that with increasing SNR age, the cut-off energy decreases, as shown in Figure \ref{fig:model_vary_age}. Equation \ref{eq:p_M} describes how the maximum energy of trapped particles decreases with age for SNR evolution during the Sedov-Taylor phase, i.e. at times $t_{\rm rad}\geq t \geq t_{\rm sed}$, as high energy particles escape. 

\begin{figure}[htp]
\centering
\includegraphics[trim=0mm 0mm 8mm 0mm,width=0.6\linewidth]{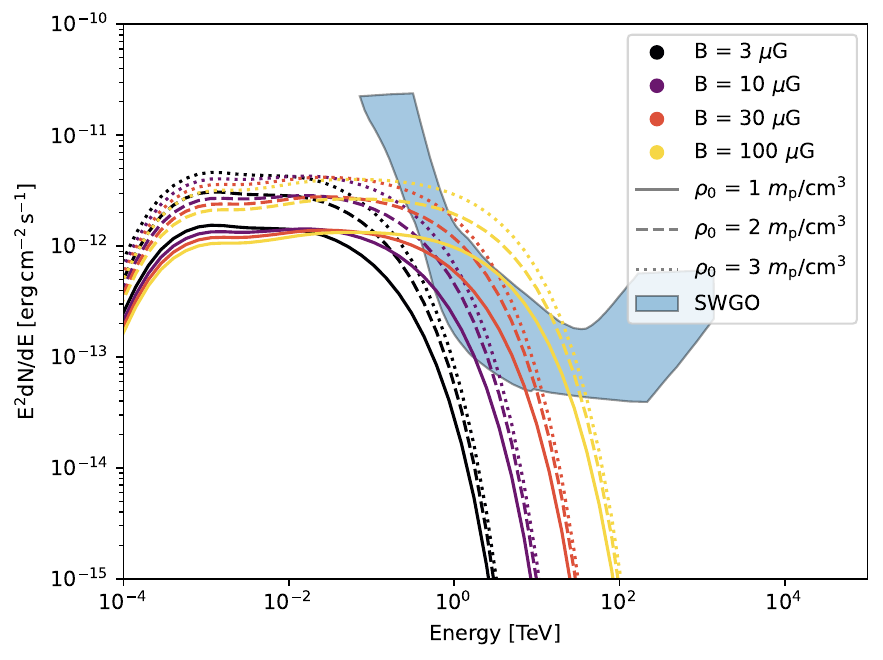}
\includegraphics[trim=0mm 0mm 8mm 0mm,width=0.6\linewidth]{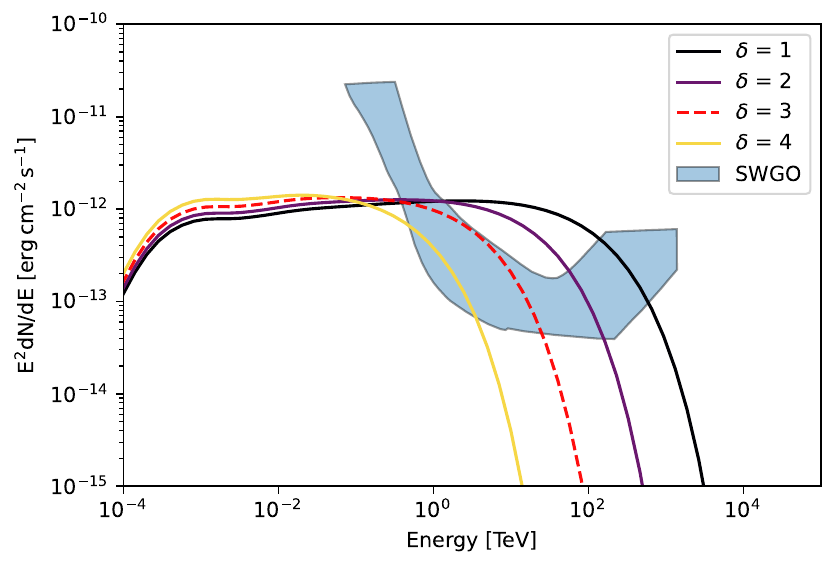}
\includegraphics[trim=0mm 0mm 8mm 0mm,width=0.6\linewidth]{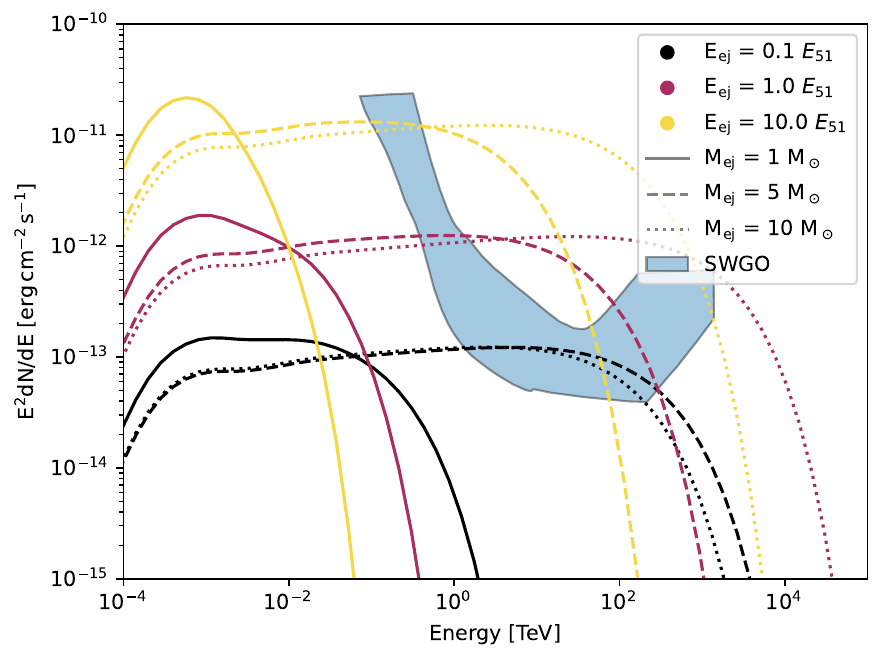}
\caption{ Influence of various model parameters on the gamma-ray SED, shown for an SNR age of 2.8\,kyr and a distance of 3.7\,kpc corresponding to the mean age and distance of the SNRs from the sample which exceed SWGO sensitivity. We take the target SWGO sensitivity band from \citep{2023arXiv230904577C} for comparison. Top: influence of magnetic field strength and ISM density on the SED. Middle: influence of $\delta$, see Equation \ref{eq:p_M}. Bottom: Influence of the SN ejecta energy $E_{\rm ej}$ and mass $M_{\rm ej}$ on the SED.}
\label{fig:model_vary}
\end{figure}

Secondly, the magnetic field strength within the SNR is treated as affecting the maximum energy achieved via Equation \ref{eq:emax}. Hence increasing the magnetic field strength also increases the maximum energy of the particle spectrum and the cut-off energy of the gamma-ray spectrum, as shown in Figure \ref{fig:model_vary}. The normalisation of the gamma-ray flux due to p-p interactions can be seen to remain at a near constant flux level below $\simeq0.1$\,TeV. 

Figure \ref{fig:model_vary} (top) also shows the influence of the ISM density $\rho_0$. This does not affect the energy reached, but rather simply number of interactions that protons undergo and hence the flux normalisation.

The parameter $\delta$ is relevant to Equation \ref{eq:p_M} describing the evolution of the particle escape in the Sedov-Taylor phase. With increasing $\delta$ values, the dependence on time is stronger, such that for the same SNR age, the maximum energy of trapped particles is decreased with increasing $\delta$. This decreases the cut-off energy of the gamma-ray spectrum as shown in Figure \ref{fig:model_vary}; changing $\delta$ from 1 to 4 reduces the maximum energy from 1500\,TeV to 7\,TeV.

With increasing ejecta mass, the Sedov time increases (Equation \ref{eqn:time_sedov_phase}), and the maximum energy at a given age also increases. However, with increasing ejecta energy, the Sedov time decreases as in Equation \ref{eqn:time_sedov_phase}, and for fixed age the maximum energy decreases (see Figure \ref{fig:model_vary}). This is a consequence of higher energy particles escaping from the remnant at earlier times.

\section{SNR selection}
\begin{figure}[htp]
\centering
\includegraphics[width=0.9\linewidth]{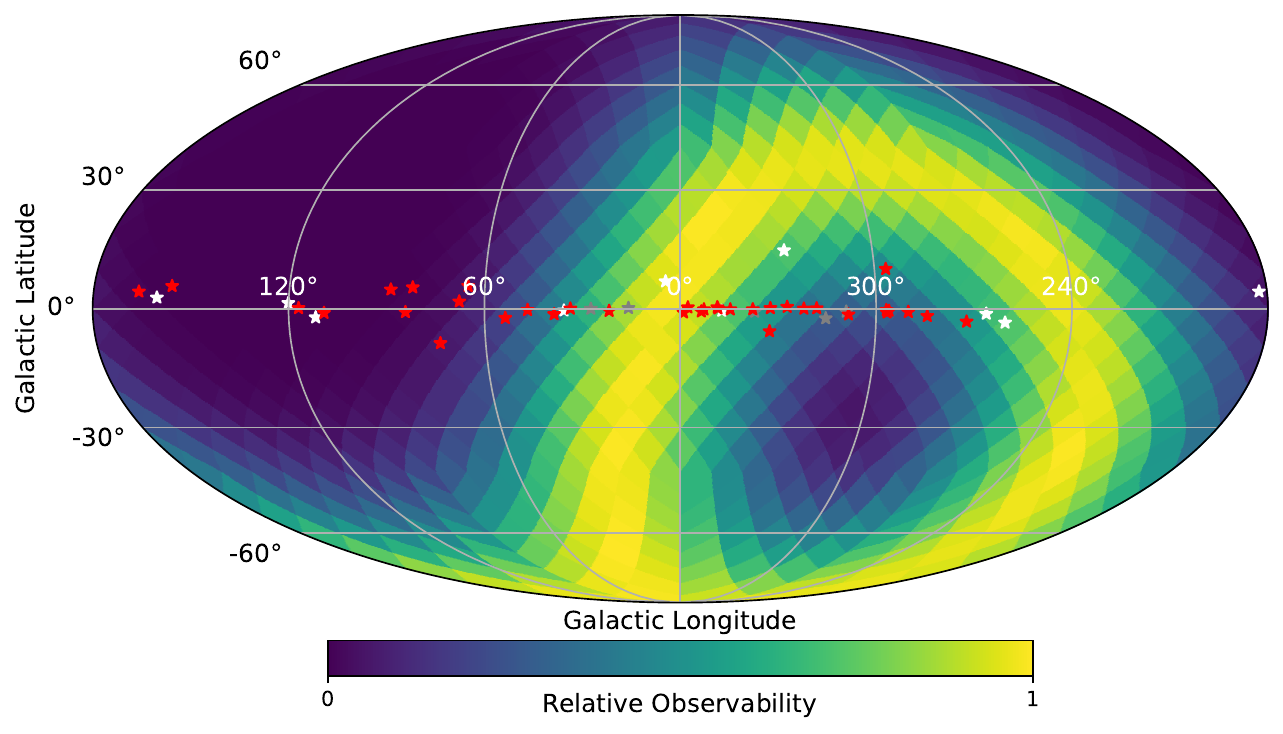}
\caption{SNRs we consider in this work. Those for which we predict a gamma-ray flux above the upper limits of the 1 year SWGO sensitivity curve shown in white, those within the band that may be detectable depending on the array configuration and analysis improvements are shown in grey, and the remaining SNRs in our selection are shown in red. Possible observability from the SWGO site relative to maximum is also shown \cite{SNRcat_article}.}
\label{fig:observability}
\end{figure}

\begin{figure}[ht!]
\centering
\includegraphics[width=0.65\linewidth]{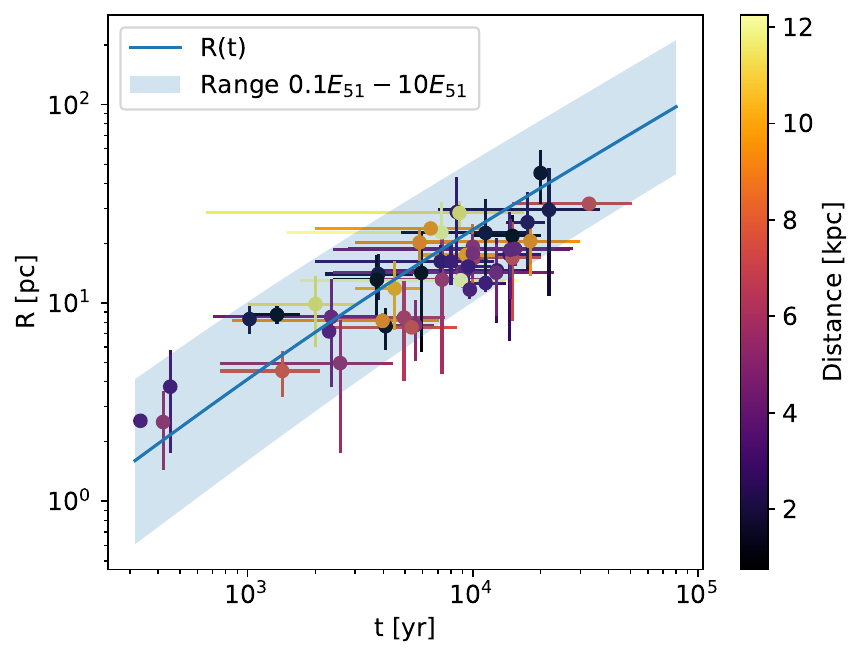}
\caption{Properties of the 55 SNRs fulfilling our selection criteria. The relation between ejecta energy $E_{\rm ej}$, age and SNR radius as described by Equation \ref{eq:radius_sedov_phase} is indicated by the linear relation, with the shaded region indicating variation in ejecta energy of a factor 10. }
\label{fig:snrcat}
\end{figure}

We obtain our set of SNRs from the public catalogue \emph{SNRcat} \cite{SNRcat_article}, which contains 383 entries of SNRs and SNR candidates. To ensure our set only contains true SNRs we exclude all entries of doubtful SNR association, such as plerionic composites and those where it is unclear whether ejecta remains. We also exclude those where information on the distance or age is not available, and restrict the age range to be $300\,\rm yr \leq \rm age \leq 95000\,\rm yr$ such that the selected SNRs are at least on the verge of the Sedov-Taylor phase. The SNRs we consider are shown in Figure \ref{fig:observability}, with the observability calculated for the SWGO site at Pampa La Bola, located at 22$^{\circ}$56’41.30“ S, 67$^{\circ}$40’39.09” W (altitude 4770\,m). A maximum observing zenith angle of $45^{\circ}$ was assumed.
Based on the reported age and radius of each SNR provided, we estimated the ejecta energy from Equation \ref{eq:radius_sedov_phase}. To down-select for SNRs that can be reasonably described by our model, we restrict the implied ejecta energy to be $0.1\,E_{51}\leq E_{\rm ej} \leq10\,E_{51}$. With these conditions we obtain the set of 51 SNRs shown in Figure \ref{fig:snrcat}. SNRs for which the measured radius and reported age lie outside of the shaded band cannot be reasonably approximated as following the Sedov-Taylor evolution described by our model. 

\begin{figure}[htp]
\centering
\includegraphics[width=0.9\linewidth]{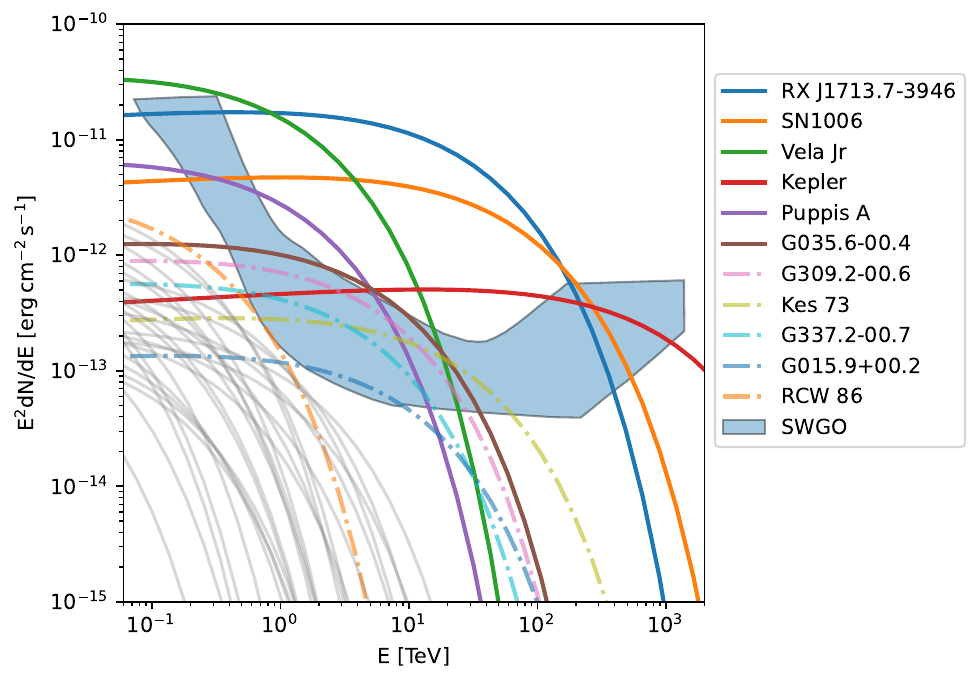}
\caption{SEDs resulting from our model for the SNRs located in the southern sky compared to the target SWGO sensitivity curve which we take from \citep{2023arXiv230904577C} for comparison. The potentially detectable SNRs are highlighted, with solid lines indicating that the SNR will be detectable regardless of array configuration and analysis method, and dot-dashed lines showing those that will be detectable in the most favorable scenario.\replaced{ A}{, a}s RCW 86 is a borderline case we will consider it potentially detectable in this work.}
\label{fig:SEDs_detect}
\end{figure}

Using Equation \ref{eq:J_p} we calculate proton spectra assuming the same conditions for each SNR, and using the GAMERA modelling package we obtain the gamma-ray curves shown in Figure \ref{fig:SEDs_detect}. 
The 6 SNRs that are located within the observable sky to SWGO and exceed the baseline SWGO sensitivity, and thus will certainly be detectable, are highlighted with solid lines in Figure \ref{fig:SEDs_detect}. Those SNRs that (according to our model) cannot be detected within 1\,year by the baseline configuration of SWGO on account of their gamma-ray flux, but that could be detected with the best possible SWGO sensitivity, are shown with dot-dashed lines in Figure \ref{fig:SEDs_detect}. We note that should improvements in the core energy range achieve the lower bound of the anticipated sensitivity band, then a further 5 SNRs are potentially detectable by SWGO in 1 year. Improvements at high energies\replaced{ (}{, }such as due to a sparse array layout\replaced{)}{,} do not lead to a larger number of detected SNRs, as few SNRs have a $E_{\mathrm{max}}\gtrapprox1\,\mathrm{TeV}$ using our model. These 11 SNRs that could potentially be detected by SWGO are listed in Table \ref{tab:SNRdect}, the other SNRs in the sample are shown in grey in Figure \ref{fig:SEDs_detect} and listed in Table \ref{tab:SNRlist}.

\begin{table}[]
\centering
\begin{tabular}{cccccc}
\hline
Source & \begin{tabular}[c]{@{}c@{}}Age [kyr]\\ $[$min, max$]$\end{tabular} & \begin{tabular}[c]{@{}c@{}}Dist. [kpc]\\ $[$min, max$]$\end{tabular} & \begin{tabular}[c]{@{}c@{}}Size\\ \replaced{[arcmin]}{}\end{tabular} & \begin{tabular}[c]{@{}c@{}}Flux\\ $E > 1\,$TeV\\ $[$TeV\,cm$^{-2}$s$^{-1}]$\end{tabular} & \begin{tabular}[c]{@{}c@{}}Flux\\ $E > 10\,$TeV\\ $[$TeV\,cm$^{-2}$s$^{-1}]$\end{tabular} \\ \hline

RX J1713.7-3946 & [1.0,1.7] & [0.9,1.1] & 60 & 3e-11 & 7.99e-12 \\
SN1006 & 1.019 & [1.6,2.2] & 30 & 1.07e-11 & 3.93e-12 \\
Vela Jr & [2.4,5.1] & [0.5,1.0] & 120 & 8.61e-12 & 8.81e-14 \\
Kepler & 0.421 & [3.3,8.2] & 3 & 1.98e-12 & 1.23e-12 \\
Puppis A & [2.2,5.4] & [1.3,2.2] & 55 & 1.52e-12 & 1.42e-14 \\
G035.6-00.4 & $\leq$2.3 & [3.5,4.1] & 13 & 1.03e-12 & 9.19e-14 \\
G309.2-00.6 & [0.7,4.0] & [2.0,7.0] & 13 & 7.08e-13 & 5.92e-14 \\
Kes 73 & [0.75,2.1] & [5.8,9.8] & 4 & 4.67e-13 & 1.15e-13 \\
G337.2-00.7 & [0.75,4.4] & [2.0,9.4] & 6 & 3.73e-13 & 2.31e-14 \\
G015.9+00.2 & [1.0,3.0] & [7.0,16.0] & 5.9 & 1.41e-13 & 1.82e-14 \\
RCW 86 & [2.0,12.4] & [2.1,3.2] & 42 & 2.98e-14 & -- \\
\hline
\end{tabular}
\caption{List of SNRs that will be potentially detectable by SWGO within 1 year according to Figure \ref{fig:SEDs_detect}. The values for the ages, distances and sizes are taken from\textit{ SNRcat }\cite{SNRcat_article}.}
\label{tab:SNRdect}
\end{table}

\section{Results for known SNRs}

To validate our simple model with existing observations, we used a Markov Chain Monte Carlo (MCMC) fitting approach for four SNRs. We fit our model to SNRs that have already been detected at gamma-ray energies, and that should also be visible from the SWGO site. From Table \ref{tab:SNRdect} (which also shows the SNRs' properties) and Figure \ref{fig:SEDs_detect}, the certainly detectable SNRs are RX\,J1713.7-3946, SN1006, Vela\,Jr, the Kepler SNR, Puppis\,A and  G035.6-00.4, and the SNRs detectable dependent on the final array configuration and analysis pipeline are G309.2-00.6, Kes\,73, G337.2-00.7 and G015.9+00.2. RCW\,86 is a borderline case which we will consider potentially detectable for the remainder of this paper. Of these, we choose four SNRs as exemplars to test the validity of our model. It should however be noted that our model is simplistic in order to maintain its predictive power, and so for interpretation of the physics of these systems in greater depth we refer the reader to the works referenced in this section.
\subsection{RX\,J1713.7-3946}

\begin{figure*}
    \centering
    \begin{minipage}{.495\textwidth}
    \centering
    \includegraphics[width=\linewidth]{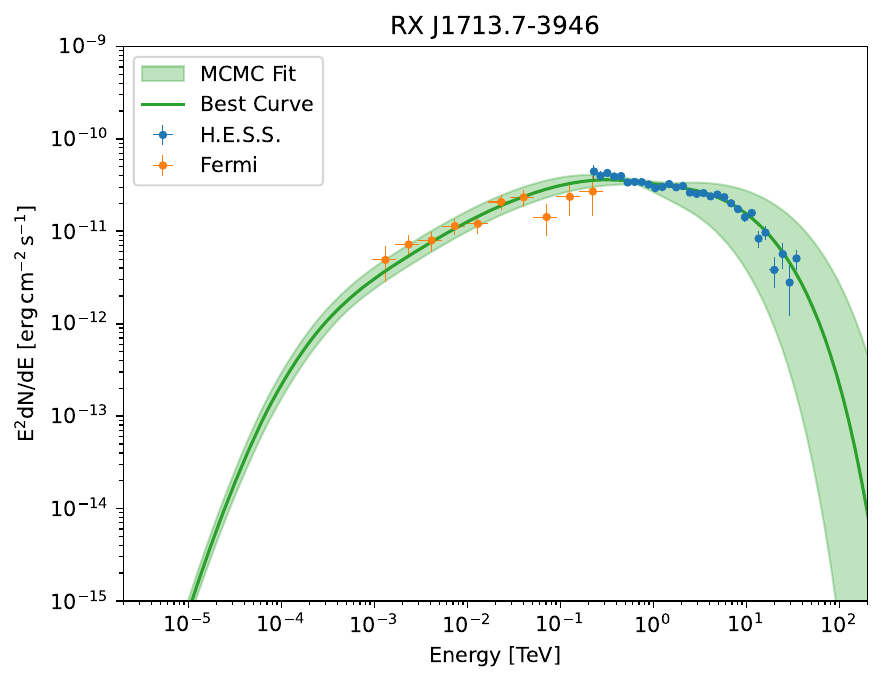}
    \end{minipage}
    \begin{minipage}{.495\textwidth}
    \centering
    \includegraphics[width=\linewidth]{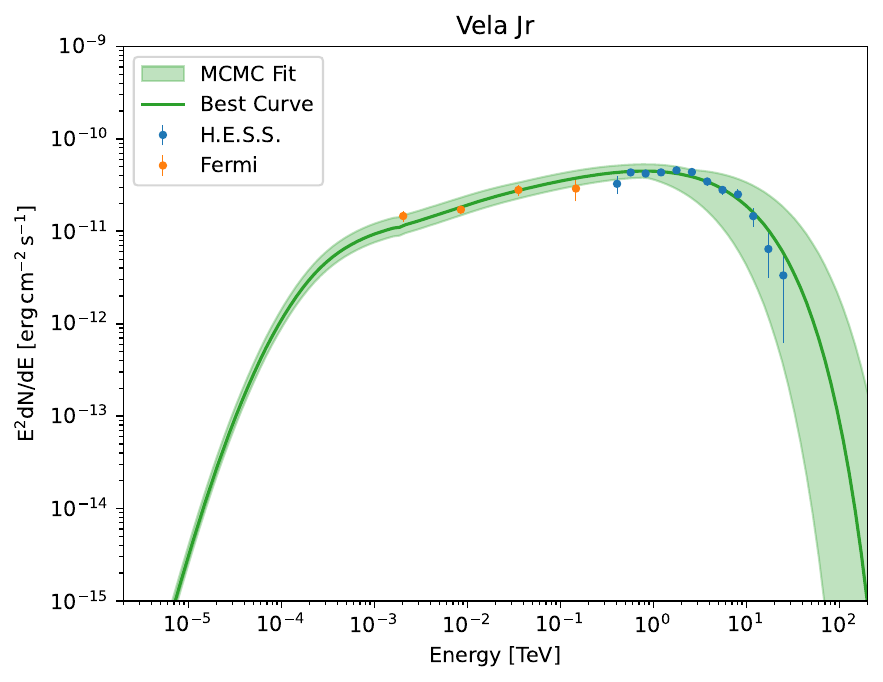}
    \end{minipage}
    \begin{minipage}{.495\textwidth}
    \includegraphics[width=\linewidth]{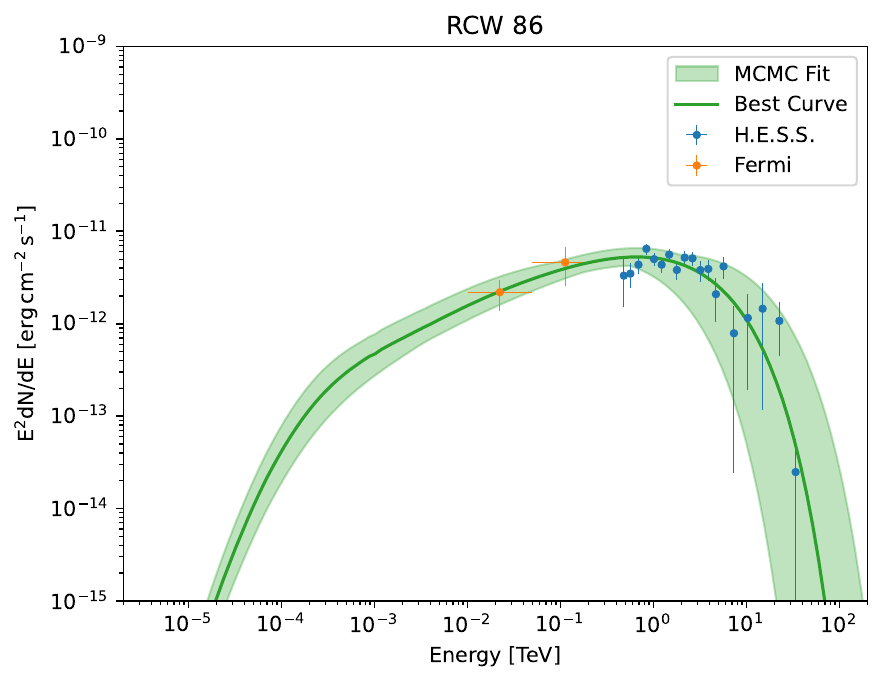}
    \end{minipage}
    \begin{minipage}{.495\textwidth}
    \includegraphics[width=\linewidth]{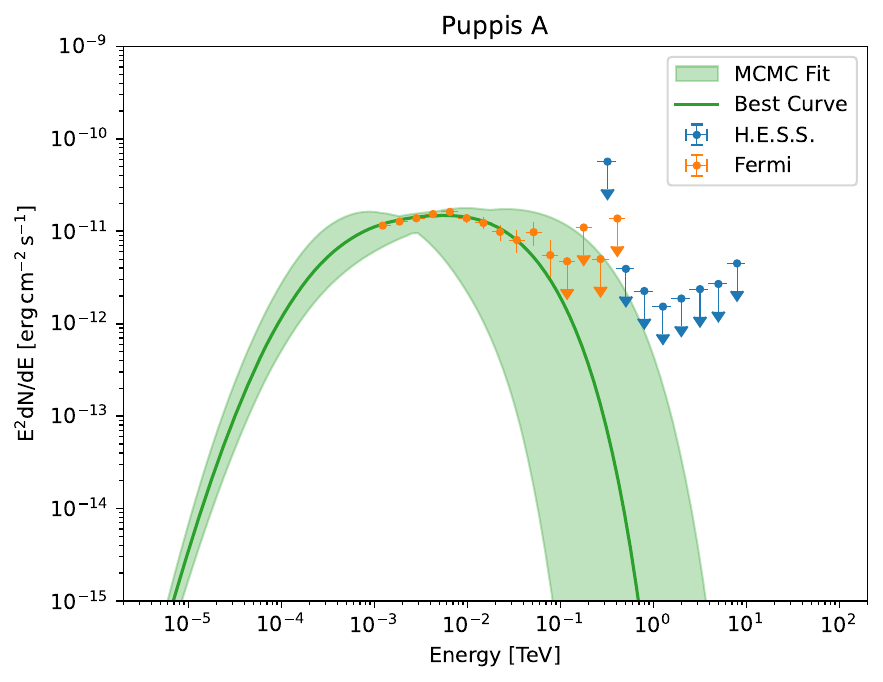}
    \end{minipage}
    \caption{Best fit results of MCMC fitting of our model to gamma-ray data for the detected SNRs. The shaded band indicates the region allowed by variation within the 16\,th and 84\,th quantiles of the sample distribution.}
    \label{fig:snr_mcmc}
\end{figure*}

For RX\,J1713.7-3946 the single power law (PL) from Equation \ref{eq:J_p} proved to not be able to sufficiently represent the data, therefore we adopted a broken power law (BPL)
\begin{equation}
    J_\mathrm{p}=
    \begin{cases}
        \left(\frac{E}{1\,\mathrm{TeV}}\right)^{-\alpha_0} & E<E_\mathrm{break}\\[1mm]
        \left(\frac{E}{E_\mathrm{break}}\right)^{-(\alpha-\alpha_0)} E^{-\alpha} \exp\left(-\frac{E}{E_\mathrm{0,p}}\right) & E\geq E_\mathrm{break}
    \end{cases}
\label{eq:BPL-HT}
\end{equation}
as used in \cite{2018_RXJ1713_HESS}. 
Figure \ref{fig:snr_mcmc} shows the results of an MCMC fit to existing gamma-ray data on this SNR,  with best fit parameters reported in Table \ref{tab:mcmc_best_values}.  

In our model, the ambient density and the ejecta energy have a similar effect on the spectrum (see Figure \ref{fig:model_vary}) and are therefore not constrained in the MCMC results. However, during the Sedov-Taylor phase they describe the SNR's radius via Equation \ref{eq:radius_sedov_phase}. Therefore, we added a condition to our prior for the MCMC fitting that checks whether the radius lies within the range of the size derived from the parameters in\textit{ SNRcat }if the age of the SNR is larger than $t_{\rm sed}$. A detailed description is given in Appendix \ref{sec:mcmc}. A similar degeneracy exists between the magnetic field and the mass in our model. Here we used evidence that the magnetic field should be at the order of $100\, \rm \mu G$ to explain the measured flux \citep{gabici_hadronic_2014}, and set the prior for $B$ to a Gaussian with $\mu=100\,\rm \mu G$ and $\sigma=20\,\rm \mu G$. For SNRs with an age of a few kyr, the influence of $\delta$ on the resulting spectrum is still weak. Therefore we set a Gaussian prior for $\delta$ as well, with $\mu=3$ and $\sigma=0.5$.

The spectral best-fit parameters differ slightly from those reported in \cite{2018_RXJ1713_HESS}, which obtained an $E_{\rm break}=1.4^{+0.7}_{-0.4}\,$TeV and $\alpha=1.94 \pm 0.05$ above the break energy, although the low index of $1.53\pm 0.09$ is compatible with our $\alpha_0=1.42\,^{+0.11}_{-0.14}$. 
In terms of physical properties, we obtained an ejecta energy of $2.75\,^{+0.12}_{-0.13}\times 10^{51}\,{\rm erg}$ for an underdense medium of $0.38\pm0.02\,{\rm cm}^{-3}$. The total energy in particles can be obtained from ejecta energy via $W_p = \varepsilon_{\rm cr}E_{\rm ej}\left(\frac{n}{1\mathrm{cm}^{-3}}\right)$ \cite{2018_RXJ1713_HESS,2018A&A...612A...7HESS_VelaJr}. For a fixed conversion efficiency $\varepsilon_{\rm cr}$ of 10\% (which we assume throughout this work), this corresponds to $\sim(1.05\pm0.07)\times10^{50}\,{\rm erg}$, a factor 10 higher than than the $10^{49}\,{\rm erg}$ reported in \cite{2018_RXJ1713_HESS}. 

Using Equations \ref{eq:p_M} and \ref{eqn:time_sedov_phase}, the cut-off energy $E_{0,\mathrm{p}}$ of the particle spectrum can be derived from our best-fit parameters in Table \ref{tab:mcmc_best_values}, yielding $60\,^{+9}_{-8}$\,TeV.

\subsection{Vela Junior}

For Vela Jr, the spectral model used for the particle spectrum was Equation \ref{eq:J_p} (consistent with that used by H.E.S.S. \cite{2018A&A...612A...7HESS_VelaJr}), where an index $\alpha_{\rm p}$ of $1.81\pm 0.08$ was fit. This is compatible with our MCMC results.

We obtained an ejecta energy for the progenitor supernova explosion of $1.18^{+0.22}_{-0.19} \times 10^{51}\,{\rm erg}$. For our obtained density of $0.71^{+0.14}_{-0.12}\,{\rm cm}^{-3}$ this corresponds to an energy in particles of $\sim (8.4 \pm 2.0) \times10^{49}\,{\rm erg}$, of the same order of magnitude as the $(7.1\pm0.3)\times10^{49}\,{\rm erg}$ reported by \cite{2018A&A...612A...7HESS_VelaJr}.

\subsection{RCW\,86} 

From our power law spectral model for the proton population, we obtain an index $\alpha_{\rm p}=1.57^{+0.13}{-0.15}$, compatible with the spectral index value of $1.7$ obtained by H.E.S.S. \cite{2018A&A...612A...4HESS_RCW86}. For RCW\,86, the corresponding cut-off energy derived from our MCMC results in Table \ref{tab:mcmc_best_values} is 30\,$^{+8}_{-7}$\,TeV. We obtain a magnetic field strength of $100^{+21}_{-20}\,\mu$G, compatible with the $\sim100\,\mu$G used for the hadronic model of \citep{2018A&A...612A...4HESS_RCW86}. We also obtain a total energy in particles of $(9.5^{+3.2}_{-2.6})\times 10^{49}\,{\rm erg}$ using our best-fit density value and a 10\% conversion efficiency from the SN ejecta energy into protons, comparable with the $\sim 7\times10^{49}\,{\rm erg}$ from \cite{2018A&A...612A...4HESS_RCW86,2009ApJ...692.1500A_RCW86}.

\subsection{Puppis\,A}

Despite a reasonably bright detection with Fermi-LAT, observations with H.E.S.S. found no significant emission, placing constraining upper limits on the region \cite{2015A&A...575A..81HESS_PuppisA}. We apply an MCMC fit to the Fermi-LAT and H.E.S.S. data with our model, finding a slightly lower ejecta mass $M_{\rm ej}=1.4^{+0.6}_{-0.5}\,M_\odot$ than for the other SNRs modelled. The cut-off energy $E_{0,\mathrm{p}}$ for the proton spectrum is however considerably lower than for the other three SNRs, which is necessary in this model to account for the lack of TeV emission from Puppis A. 

\begin{table}[]
\centering
\begin{tabular}{ccccc}
\hline
 & RX J1713.7-3946 & Vela Jr & RCW 86 & Puppis A \\ \hline
$J_p$ & BPL & PL & PL & PL \\
$B$ ($\mu$G) & 100\,$\pm$\,20 & 99\,$^{+20}_{-21}$ & 100\,$^{+21}_{-20}$ & 99\,$\pm$\,20 \\[.5mm]
$M_{\rm ej}$ ($M_\odot$) & 2.7\,$^{+0.5}_{-0.6}$ & 4.3\,$^{+0.8}_{-0.9}$ & 4.8\,$^{+1.0}_{-1.1}$ & 1.4\,$^{+0.6}_{-0.5}$ \\[.5mm]
$\rho$ ($m_{\rm p}$cm$^{-3}$) & 0.38\,$\pm$\,0.02 & 0.71\,$^{+0.14}_{-0.12}$ & 1.44\,$^{+0.40}_{-0.29}$ & 1.12\,$^{+0.12}_{-0.11}$ \\[.5mm]
$\delta$ & 3.0\,$\pm$\,0.5 & 3.0\,$\pm$\,0.5 & 3.0\,$\pm$\,0.5 & 3.1\,$\pm$\,0.5 \\[.5mm]
$E_{\rm ej}$ ($E_{51}$) & 2.75\,$^{+0.12}_{-0.13}$ & 1.18\,$^{+0.22}_{-0.19}$ & 0.66\,$^{+0.13}_{-0.12}$ & 1.36\,$^{+0.10}_{-0.09}$ \\[.5mm]
$\alpha_{\rm p}$ & -- & 1.78\,$\pm$\,0.03 & 1.57\,$^{+0.13}_{-0.15}$ & 1.42\,$^{+0.23}_{-0.40}$ \\[.5mm]
$\beta_{\rm p}$ & -- & 1.25\,$^{+0.22}_{0.21}$ & 1.27\,$^{+0.21}_{-0.20}$ & 0.95\,$^{+0.26}_{-0.25}$ \\[.5mm]
$\alpha$ & 1.54\,$^{+0.3}_{-0.4}$ & -- & -- & -- \\[.5mm]
$\alpha_0$ & 1.42\,$^{+0.11}_{-0.14}$ & -- & -- & -- \\[.5mm]
$E_{\rm break}$ (TeV) & 3.2\,$^{+0.9}_{-0.7}$ & -- & -- & -- \\[.5mm]
$E_{\rm 0,p}$ (TeV) & 61\,$^{+150}_{-21}$ & 66\,$^{+160}_{-23}$ & 28\,$^{+80}_{-9}$ & 0.132\,$^{+0.700}_{-0.016}$\\
\hline
\end{tabular}
\caption{Best fit values from the MCMC, the values correspond to the 16th, 50th and 84th quantile of the sample distribution. Corner and convergence plots are shown in Appendix \ref{sec:mcmc}.}
\label{tab:mcmc_best_values}
\end{table}

\section{Discussion and conclusions}

In this study, we implemented a simple evolutionary model to describe SNRs in the Sedov phase, during which it is most likely that gamma-ray emission can be anticipated. Using \textit{SNRcat} \cite{SNRcat_article}, we selected SNRs that will be observable by the forthcoming SWGO. Implementing the baseline scenario for our model (representative values) for each SNR, and adapting the age and distance as appropriate, we are able to predict an expected gamma-ray flux level. A total of at least six and possibly as many as eleven SNRs have predicted gamma-ray fluxes that should be detectable by SWGO within one year of operation.

For four of these SNRs that have already been detected at gamma-ray energies, RX\,J1713.7-3946, Vela\,Jr, RCW\,86 and Puppis A, we proceeded to run an MCMC analysis using our model to constrain the properties of the supernova remnant and its environment. These converged to values that are comparable to those in the literature, as discussion in the previous section. 

This broad consistency with values obtained in previous works helps to validate our simple model and predictions for SNR observations with SWGO. Given that WCD facilities such as SWGO have a wide field-of-view, observations may be able to detect a halo of particles escaped from the SNR shock, as predicted by \cite{2021A&A...654A.139Brose}, that may extend over large angular scales. Detailed studies of large-scale gamma-ray emission around SNRs could constrain particle transport and the diffusion coefficient of CRs in the vicinity of SNR shocks, as well as the contribution of escaping particles to the galactic CR sea. 

Although we do not fit \deleted{the }SN\,1006 with our model in this study, observations of the SNR will nevertheless be interesting\deleted{,} on account of its dual lobe morphology. It has been suggested that this feature is due to polarisation of the magnetic field in this SNR, such that particles are preferentially transported in one direction along the magnetic field lines, as opposed to perpendicular to them (where less emission is observed) \cite{2023ApJ...957...55ZhouSN1006}. Observations with SWGO could further help to corroborate this hypothesis, by improving the sensitivity across the entire region and on larger angular scales, indicating to what extent the nonthermal emission in the region continues to show preferential orientation. 

SWGO will have an angular resolution improving with energy from $\sim1^\circ$ at threshold to $\lesssim0.4^\circ$ at energies $>1$\,TeV, potentially reaching $\sim 0.1^\circ$ \cite{SWGO_white_paper}. From Table \ref{tab:SNRdect}, this should be sufficient to resolve shell-like structure in at least half of the potentially detectable SNRs. Asymmetries to the emission could hence be detected by SWGO, although these are not accounted for in our \replaced{work}{model}.  

The model presented here can be similarly applied to \replaced{SNRs}{supernova remnants} located in the Northern hemisphere, and potentially compared to Northern observations by facilities such as HAWC, LHAASO,\added{ the ASTRI Mini-Array} or the forthcoming CTA\added{O} \citep{hawc,lhaaso,Scuderi_2022,CTA_book}. However, to gain a better handle on the magnetic field strength or ambient density, multiwavelength observations are required. Given the large angular size of some of these supernova remnants, sky scanning facilities such as eROSITA \citep{2021A&A...647A...1Predehl_eROSITA} or MeerKAT \citep{goedhart2024saraomeerkat13ghz} provide ideal complements to the SWGO survey view at TeV energies. 

Recent results from HAWC and LHAASO revealed a new population of ultra high energy ($E_\gamma>100\,$TeV) sources along the Northern galactic plane \citep{HAWC_56tev,2021Cao_uhe,Cao_1lhaaso}. We anticipate that a number of ultra high energy sources remain to be discovered in the Southern sky by SWGO. Whether or not SNRs are among these ultra high energy sources remains to be determined.  

As we assumed a hadronic scenario in all cases, constraining the model parameters\replaced{ (}{, }for both the SN energetics and proton spectrum\replaced{)}{,} yields insights as to the likely contributions of the SNRs towards the origins of galactic Cosmic Rays. It should also be noted that SWGO will also have a high degree of complementarity with the upcoming Cherenkov Telescope Array Observatory (CTAO); new SNRs detected by SWGO given its wide field of view could be the subject of follow-up observations by CTAO with its enhanced angular resolution. This would help to reduce source confusion. But SWGO may be sensitive to larger angular structures, and will have a larger effective area at the highest energies, therefore constraining the maximum energies reached.

\acknowledgments


NS, STS and AMWM are supported by the Deutsche Forschungsgemeinschaft (DFG) project number 452934793. 



\bibliographystyle{JHEP}
\bibliography{biblio.bib}

\providecommand{\href}[2]{#2}\begingroup\raggedright\begin{thebibliography}{10}

\bibitem{Hinton_2009}
J.~Hinton and W.~Hofmann, \emph{Teraelectronvolt astronomy},
  \href{https://doi.org/10.1146/annurev-astro-082708-101816}{\emph{Annual
  Review of Astronomy and Astrophysics} {\bfseries 47} (2009) 523–565}.

\bibitem{2023ApJ...958....3D}
R.~{Diesing}, \emph{{The Maximum Energy of Shock-accelerated Cosmic Rays}},
  \href{https://doi.org/10.3847/1538-4357/ad00b1}{\emph{apj} {\bfseries 958}
  (2023) 3} [\href{https://arxiv.org/abs/2305.07697}{{\ttfamily 2305.07697}}].

\bibitem{christofari}
{Cristofari, P.}, {Blasi, P.} and {Caprioli, D.}, \emph{Cosmic ray protons and
  electrons from supernova remnants},
  \href{https://doi.org/10.1051/0004-6361/202140448}{\emph{A\&A} {\bfseries
  650} (2021) A62}.

\bibitem{aharoniangamma}
F.A.~{Aharonian}, \emph{{Gamma rays from supernova remnants}},
  \href{https://doi.org/10.1016/j.astropartphys.2012.08.007}{\emph{Astroparticle
  Physics} {\bfseries 43} (2013) 71}.

\bibitem{2023arXiv230904577C}
R.~{Concei{\c{c}}{\~a}o}, \emph{{The Southern Wide-field Gamma-ray
  Observatory}}, \href{https://doi.org/10.48550/arXiv.2309.04577}{\emph{arXiv
  e-prints} (2023) arXiv:2309.04577}
  [\href{https://arxiv.org/abs/2309.04577}{{\ttfamily 2309.04577}}].

\bibitem{2015ICRC...34..917Hahn}
J.~{Hahn}, \emph{{GAMERA - a new modeling package for non-thermal spectral
  modeling}},  in \emph{34th International Cosmic Ray Conference (ICRC2015)},
  vol.~34 of \emph{International Cosmic Ray Conference}, p.~917, July, 2015,
  \href{https://doi.org/10.22323/1.236.0917}{DOI}.

\bibitem{2022ascl.soft03007HahnGAMERA}
J.~{Hahn}, C.~{Romoli} and M.~{Breuhaus}, ``{GAMERA: Source modeling in gamma
  astronomy}.'' Astrophysics Source Code Library, record ascl:2203.007, Mar.,
  2022.

\bibitem{SupernovaRemnantsWoltjer}
L.~Woltjer, \emph{Supernova remnants},
  \href{https://doi.org/10.1146/annurev.aa.10.090172.001021}{\emph{Annual
  Review of Astronomy and Astrophysics} {\bfseries 10} (1972) 129}
  [\href{https://arxiv.org/abs/https://doi.org/10.1146/annurev.aa.10.090172.001021}{{\ttfamily
  https://doi.org/10.1146/annurev.aa.10.090172.001021}}].

\bibitem{CARDILLO20151}
M.~Cardillo, E.~Amato and P.~Blasi, \emph{On the cosmic ray spectrum from type
  ii supernovae expanding in their red giant presupernova wind},
  \href{https://doi.org/https://doi.org/10.1016/j.astropartphys.2015.03.002}{\emph{Astroparticle
  Physics} {\bfseries 69} (2015) 1}.

\bibitem{1999_Truelove_McKee}
J.K.~{Truelove} and C.F.~{McKee}, \emph{{Evolution of Nonradiative Supernova
  Remnants}}, \href{https://doi.org/10.1086/313176}{\emph{apjs} {\bfseries 120}
  (1999) 299}.

\bibitem{vink2020physics}
J.~Vink, \emph{Physics and evolution of supernova remnants}, Springer (2020).

\bibitem{1978MNRASBell}
A.R.~{Bell}, \emph{{The acceleration of cosmic rays in shock fronts - I.}},
  \href{https://doi.org/10.1093/mnras/182.2.147}{\emph{MNRAS} {\bfseries 182}
  (1978) 147}.

\bibitem{Blasi_2013}
P.~Blasi, \emph{The origin of galactic cosmic rays},
  \href{https://doi.org/10.1007/s00159-013-0070-7}{\emph{The Astronomy and
  Astrophysics Review} {\bfseries 21} (2013) }.

\bibitem{Celli_2020}
S.~{Celli}, F.~{Aharonian} and S.~{Gabici}, \emph{{Spectral Signatures of
  PeVatrons}}, \href{https://doi.org/10.3847/1538-4357/abb805}{\emph{apj}
  {\bfseries 903} (2020) 61}
  [\href{https://arxiv.org/abs/2009.05999}{{\ttfamily 2009.05999}}].

\bibitem{Celli_thesis}
S.~Celli, \emph{Gamma-ray and Neutrino Signatures of Galactic Cosmic-ray
  Accelerators}, Springer, Cham, CH (2019),
  \href{https://doi.org/https://doi.org/10.1007/978-3-030-33124-5}{https://doi.org/10.1007/978-3-030-33124-5}.

\bibitem{gabici09}
S.~Gabici, F.A.~Aharonian and S.~Casanova, \emph{Broad–band non–thermal
  emission from molecular clouds illuminated by cosmic rays from nearby
  supernova remnants},
  \href{https://doi.org/10.1111/j.1365-2966.2009.14832.x}{\emph{Monthly Notices
  of the Royal Astronomical Society} {\bfseries 396} (2009) 1629}
  [\href{https://arxiv.org/abs/https://academic.oup.com/mnras/article-pdf/396/3/1629/5800874/mnras0396-1629.pdf}{{\ttfamily
  https://academic.oup.com/mnras/article-pdf/396/3/1629/5800874/mnras0396-1629.pdf}}].

\bibitem{2014PhRvD..90l3014Kafexhiu}
E.~{Kafexhiu}, F.~{Aharonian}, A.M.~{Taylor} and G.S.~{Vila},
  \emph{{Parametrization of gamma-ray production cross sections for p p
  interactions in a broad proton energy range from the kinematic threshold to
  PeV energies}}, \href{https://doi.org/10.1103/PhysRevD.90.123014}{\emph{prd}
  {\bfseries 90} (2014) 123014}
  [\href{https://arxiv.org/abs/1406.7369}{{\ttfamily 1406.7369}}].

\bibitem{annurev:/content/journals/10.1146/annurev.astro.46.060407.145237}
S.P.~Reynolds, \emph{Supernova remnants at high energy},
  \href{https://doi.org/https://doi.org/10.1146/annurev.astro.46.060407.145237}{\emph{Annual
  Review of Astronomy and Astrophysics} {\bfseries 46} (2008) 89}.

\bibitem{Bamba_2022}
A.~Bamba and B.J.~Williams, \emph{Supernova remnants: Types and evolution},  in
  \emph{Handbook of X-ray and Gamma-ray Astrophysics}, p.~1–12, Springer
  Nature Singapore (2022),
  \href{https://doi.org/10.1007/978-981-16-4544-0_88-1}{DOI}.

\bibitem{diesing}
R.~{Diesing}, M.~{Guo}, C.-G.~{Kim}, J.~{Stone} and D.~{Caprioli},
  \emph{{Nonthermal Signatures of Radiative Supernova Remnants}},
  \href{https://doi.org/10.3847/1538-4357/ad74f0}{\emph{apj} {\bfseries 974}
  (2024) 201} [\href{https://arxiv.org/abs/2404.15396}{{\ttfamily
  2404.15396}}].

\bibitem{SNRcat_article}
G.~{Ferrand} and S.~{Safi-Harb}, \emph{{A census of high-energy observations of
  Galactic supernova remnants}},
  \href{https://doi.org/10.1016/j.asr.2012.02.004}{\emph{Advances in Space
  Research} {\bfseries 49} (2012) 1313}
  [\href{https://arxiv.org/abs/1202.0245}{{\ttfamily 1202.0245}}].

\bibitem{2018_RXJ1713_HESS}
{H.~E.~S.~S. Collaboration}, H.~{Abdalla}, A.~{Abramowski}, F.~{Aharonian},
  F.~{Ait Benkhali}, A.G.~{Akhperjanian} et~al., \emph{{H.E.S.S. observations
  of RX J1713.7-3946 with improved angular and spectral resolution: Evidence
  for gamma-ray emission extending beyond the X-ray emitting shell}},
  \href{https://doi.org/10.1051/0004-6361/201629790}{\emph{aap} {\bfseries 612}
  (2018) A6} [\href{https://arxiv.org/abs/1609.08671}{{\ttfamily 1609.08671}}].

\bibitem{gabici_hadronic_2014}
S.~Gabici and F.A.~Aharonian, \emph{Hadronic gamma-rays from {RX}
  {J1713}.7-3946?}, \href{https://doi.org/10.1093/mnrasl/slu132}{\emph{Monthly
  Notices of the Royal Astronomical Society} {\bfseries 445} (2014) L70}.

\bibitem{2018A&A...612A...7HESS_VelaJr}
{H.~E.~S.~S. Collaboration}, H.~{Abdalla}, A.~{Abramowski}, F.~{Aharonian},
  F.~{Ait Benkhali}, A.G.~{Akhperjanian} et~al., \emph{{Deeper H.E.S.S.
  observations of Vela Junior (RX J0852.0-4622): Morphology studies and
  resolved spectroscopy}},
  \href{https://doi.org/10.1051/0004-6361/201630002}{\emph{aap} {\bfseries 612}
  (2018) A7} [\href{https://arxiv.org/abs/1611.01863}{{\ttfamily 1611.01863}}].

\bibitem{2018A&A...612A...4HESS_RCW86}
{H.~E.~S.~S. Collaboration}, A.~{Abramowski}, F.~{Aharonian}, F.~{Ait
  Benkhali}, A.G.~{Akhperjanian}, E.O.~{Ang{\"u}ner} et~al., \emph{{Detailed
  spectral and morphological analysis of the shell type supernova remnant RCW
  86}}, \href{https://doi.org/10.1051/0004-6361/201526545}{\emph{aap}
  {\bfseries 612} (2018) A4}
  [\href{https://arxiv.org/abs/1601.04461}{{\ttfamily 1601.04461}}].

\bibitem{2009ApJ...692.1500A_RCW86}
F.~{Aharonian}, A.G.~{Akhperjanian}, U.B.~{de Almeida}, A.R.~{Bazer-Bachi},
  B.~{Behera}, M.~{Beilicke} et~al., \emph{{Discovery of Gamma-Ray Emission
  From the Shell-Type Supernova Remnant RCW 86 With Hess}},
  \href{https://doi.org/10.1088/0004-637X/692/2/1500}{\emph{apj} {\bfseries
  692} (2009) 1500} [\href{https://arxiv.org/abs/0810.2689}{{\ttfamily
  0810.2689}}].

\bibitem{2015A&A...575A..81HESS_PuppisA}
{H.~E.~S.~S. Collaboration}, A.~{Abramowski}, F.~{Aharonian}, F.~{Ait
  Benkhali}, A.G.~{Akhperjanian}, E.O.~{Ang{\"u}ner} et~al., \emph{{H.E.S.S.
  reveals a lack of TeV emission from the supernova remnant Puppis A}},
  \href{https://doi.org/10.1051/0004-6361/201424805}{\emph{aap} {\bfseries 575}
  (2015) A81} [\href{https://arxiv.org/abs/1412.6997}{{\ttfamily 1412.6997}}].

\bibitem{2021A&A...654A.139Brose}
R.~{Brose}, M.~{Pohl} and I.~{Sushch}, \emph{{Morphology of supernova remnants
  and their halos}},
  \href{https://doi.org/10.1051/0004-6361/202141194}{\emph{aap} {\bfseries 654}
  (2021) A139} [\href{https://arxiv.org/abs/2108.10773}{{\ttfamily
  2108.10773}}].

\bibitem{2023ApJ...957...55ZhouSN1006}
P.~{Zhou}, D.~{Prokhorov}, R.~{Ferrazzoli}, Y.-J.~{Yang}, P.~{Slane}, J.~{Vink}
  et~al., \emph{{Magnetic Structures and Turbulence in SN 1006 Revealed with
  Imaging X-Ray Polarimetry}},
  \href{https://doi.org/10.3847/1538-4357/acf3e6}{\emph{apj} {\bfseries 957}
  (2023) 55} [\href{https://arxiv.org/abs/2309.01879}{{\ttfamily 2309.01879}}].

\bibitem{SWGO_white_paper}
A.~{Albert}, R.~{Alfaro}, H.~{Ashkar}, C.~{Alvarez}, J.~{{\'A}lvarez},
  J.C.~{Arteaga-Vel{\'a}zquez} et~al., \emph{{Science Case for a Wide
  Field-of-View Very-High-Energy Gamma-Ray Observatory in the Southern
  Hemisphere}}, \href{https://doi.org/10.48550/arXiv.1902.08429}{\emph{arXiv
  e-prints} (2019) arXiv:1902.08429}
  [\href{https://arxiv.org/abs/1902.08429}{{\ttfamily 1902.08429}}].

\bibitem{hawc}
A.U.~Abeysekara, A.~Albert, R.~Alfaro, C.~Alvarez, J.D.~Álvarez, R.~Arceo
  et~al., \emph{Observation of the crab nebula with the hawc gamma-ray
  observatory}, \href{https://doi.org/10.3847/1538-4357/aa7555}{\emph{The
  Astrophysical Journal} {\bfseries 843} (2017) 39}.

\bibitem{lhaaso}
Z.~Cao, F.~Aharonian, Q.~An, Axikegu, L.X.~Bai, Y.X.~Bai et~al.,
  \emph{Peta–electron volt gamma-ray emission from the crab nebula},
  \href{https://doi.org/10.1126/science.abg5137}{\emph{Science} {\bfseries 373}
  (2021) 425–430}.

\bibitem{Scuderi_2022}
S.~Scuderi, A.~Giuliani, G.~Pareschi, G.~Tosti, O.~Catalano, E.~Amato et~al.,
  \emph{The astri mini-array of cherenkov telescopes at the observatorio del
  teide}, \href{https://doi.org/10.1016/j.jheap.2022.05.001}{\emph{Journal of
  High Energy Astrophysics} {\bfseries 35} (2022) 52–68}.

\bibitem{CTA_book}
C.~Consortium, \emph{Science with the Cherenkov Telescope Array}, WORLD
  SCIENTIFIC (2019), \href{https://doi.org/10.1142/10986}{10.1142/10986},
  [\href{https://arxiv.org/abs/https://www.worldscientific.com/doi/pdf/10.1142/10986}{{\ttfamily
  https://www.worldscientific.com/doi/pdf/10.1142/10986}}].

\bibitem{2021A&A...647A...1Predehl_eROSITA}
P.~{Predehl}, R.~{Andritschke}, V.~{Arefiev}, V.~{Babyshkin}, O.~{Batanov},
  W.~{Becker} et~al., \emph{{The eROSITA X-ray telescope on SRG}},
  \href{https://doi.org/10.1051/0004-6361/202039313}{\emph{A\&A} {\bfseries
  647} (2021) A1} [\href{https://arxiv.org/abs/2010.03477}{{\ttfamily
  2010.03477}}].

\bibitem{goedhart2024saraomeerkat13ghz}
S.~Goedhart, W.D.~Cotton, F.~Camilo, M.A.~Thompson, G.~Umana, M.~Bietenholz
  et~al., \emph{The sarao meerkat 1.3 ghz galactic plane survey},  2024.

\bibitem{HAWC_56tev}
A.U.~{Abeysekara}, A.~{Albert}, R.~{Alfaro}, J.R.~{Angeles Camacho},
  J.C.~{Arteaga-Vel{\'a}zquez}, K.P.~{Arunbabu} et~al., \emph{{Multiple
  Galactic Sources with Emission Above 56 TeV Detected by HAWC}},
  \href{https://doi.org/10.1103/PhysRevLett.124.021102}{\emph{PRL} {\bfseries
  124} (2020) 021102} [\href{https://arxiv.org/abs/1909.08609}{{\ttfamily
  1909.08609}}].

\bibitem{2021Cao_uhe}
Z.~{Cao}, F.A.~{Aharonian}, Q.~{An}, L.X.~{Axikegu}, Bai, Y.X.~{Bai},
  Y.W.~{Bao} et~al., \emph{{Ultrahigh-energy photons up to 1.4
  petaelectronvolts from 12 {\ensuremath{\gamma}}-ray Galactic sources}},
  \href{https://doi.org/10.1038/s41586-021-03498-z}{\emph{Nature} {\bfseries
  594} (2021) 33}.

\bibitem{Cao_1lhaaso}
Z.~{Cao}, F.~{Aharonian}, Q.~{An}, {Axikegu}, Y.X.~{Bai}, Y.W.~{Bao} et~al.,
  \emph{{The First LHAASO Catalog of Gamma-Ray Sources}},
  \href{https://doi.org/10.3847/1538-4365/acfd29}{\emph{ApJS} {\bfseries 271}
  (2024) 25} [\href{https://arxiv.org/abs/2305.17030}{{\ttfamily 2305.17030}}].

\bibitem{Hogg_2018}
D.W.~Hogg and D.~Foreman-Mackey, \emph{Data analysis recipes: Using markov
  chain monte carlo*},
  \href{https://doi.org/10.3847/1538-4365/aab76e}{\emph{The Astrophysical
  Journal Supplement Series} {\bfseries 236} (2018) 11}.

\bibitem{2020SciPy-NMeth}
P.~Virtanen, R.~Gommers, T.E.~Oliphant, M.~Haberland, T.~Reddy, D.~Cournapeau
  et~al., \emph{{{SciPy} 1.0: Fundamental Algorithms for Scientific Computing
  in Python}}, \href{https://doi.org/10.1038/s41592-019-0686-2}{\emph{Nature
  Methods} {\bfseries 17} (2020) 261}.

\bibitem{Foreman_Mackey_2013}
D.~Foreman-Mackey, D.W.~Hogg, D.~Lang and J.~Goodman, \emph{\texttt{emcee}: The
  mcmc hammer}, \href{https://doi.org/10.1086/670067}{\emph{Publications of the
  Astronomical Society of the Pacific} {\bfseries 125} (2013) 306–312}.

\end{thebibliography}\endgroup

\appendix 
\section{Flux predictions for SNRs}
\label{sec:SNRtables}

\begin{table}[]
\centering
\begin{tabular}{ccccccc}
Source & Observable & \begin{tabular}[c]{@{}c@{}}Age [kyr]\\ $[$min, max$]$\end{tabular} & \begin{tabular}[c]{@{}c@{}}Dist. [kpc]\\ $[$min, max$]$\end{tabular} & \begin{tabular}[c]{@{}c@{}}Size\\ (arcmin)\end{tabular} & \begin{tabular}[c]{@{}c@{}}Flux\\ $E > 1\,$TeV\\ $[$TeV\,cm$^{-2}$s$^{-1}]$\end{tabular} & \begin{tabular}[c]{@{}c@{}}Flux\\ $E > 10\,$TeV\\ $[$TeV\,cm$^{-2}$s$^{-1}]$\end{tabular} \\ \hline
Tycho & -- & 0.453 & [1.5,5.0] & 8 & 6.02e-12 & 3.66e-12 \\
Cassiopeia A & -- & [0.316,0.352] & [3.3,3.7] & 5 & 5.13e-12 & 3.09e-12 \\
G182.4+04.3 & -- & [3.8,4.4] & [0.8,1.3] & 50 & 3.3e-12 & 1.81e-14 \\
HB9 & -- & [2.6,9.2] & [0.3,1.2] & 130 & 1.33e-12 & 1.39e-16 \\
Circinus X-1 & \checkmark & [0.85,7.1] & [8.4,10.2] & 6 & 4.66e-14 & 3.21e-16 \\
G308.3-01.4 & \checkmark & [2.4,7.5] & [3.1,9.8] & 9 & 4.25e-14 & 4.24e-17 \\
Kes 79 & \checkmark & [4.4,6.7] & [3.5,7.1] & 10 & 3.68e-14 & 9.29e-18 \\
G344.7-00.1 & \checkmark & [3.0,6.0] & [6.3,14.0] & 8 & 2.53e-14 & 6.43e-17 \\
G352.7-00.1 & \checkmark & [2.2,8.5] & [7.0,8.0] & 6.9 & 2.2e-14 & 8.94e-18 \\
G318.2+00.1 & \checkmark & $\leq$8.0 & [2.3,3.71] & 37 & 9.97e-15 & -- \\
Kes 32 & \checkmark & [3.0,8.6] & [7.5,11.0] & 15 & 9.59e-15 & 1.3e-18 \\
PKS 1209-51/52 & \checkmark & [7.0,10.0] & [1.3,3.9] & 76 & 7.67e-15 & -- \\
G272.2-03.2 & \checkmark & [3.6,11.0] & [2.0,10.0] & 15 & 5.25e-15 & -- \\
1156-62 & \checkmark & [2.0,11.0] & [9.0,10.2] & 17 & 4.58e-15 & -- \\
G332.5-05.6 & \checkmark & [7.0,12.1] & [2.2,3.8] & 35 & 1.7e-15 & -- \\
G067.7+01.8 & -- & [1.5,13.0] & [7.0,17.0] & 13 & 1.38e-15 & -- \\
G038.7-01.3 & \checkmark & [3.8,14.7] & [4.0,4.0] & 25 & 1.37e-15 & -- \\
Kes 69 & \checkmark & [8.8,9.2] & [4.1,5.8] & 20 & 1.2e-15 & -- \\
G359.0-00.9 & \checkmark & $\leq$9.7 & [3.13,3.85] & 23 & 1.05e-15 & -- \\
HB21 & -- & [4.8,18.0] & [0.8,2.3] & 100 & 5.91e-16 & -- \\
the Square & \checkmark & $\leq$10.0 & [3.79,6.4] & 24 & 3.39e-16 & -- \\
CTB 37B & \checkmark & [0.65,16.8] & [9.8,13.2] & 17 & 3.04e-16 & -- \\
MSH 10-53 & \checkmark & $\leq$10.0 & [4.8,6.2] & 24 & 2.91e-16 & -- \\
Kes 41 & \checkmark & [1.7,16.0] & [12.2,12.3] & 7.3 & 2.32e-16 & -- \\
G084.2-00.8 & -- & [8.4,11.7] & [5.8,7.0] & 18 & 2.02e-16 & -- \\
G296.7-00.9 & \checkmark & [5.8,12.9] & [9.1,10.9] & 12 & 1.94e-16 & -- \\
CTB 109 & -- & [8.8,14.0] & [2.79,3.4] & 28 & 1.48e-16 & -- \\
HC30 & \checkmark & $\leq$10.0 & [5.7,11.4] & 15 & 1.2e-16 & -- \\
CTB 1 & -- & [7.5,18.1] & [1.6,4.3] & 34 & 2.14e-17 & -- \\
Kes 27 & \checkmark & [2.4,23.0] & [2.8,6.5] & 21 & 1e-17 & -- \\
Cygnus Loop & -- & [10.0,20.0] & [0.576,1.0] & 190 & 8.08e-18 & -- \\
VRO 42.05.01 & -- & [9.0,20.1] & [1.0,4.5] & 44 & 1.46e-18 & -- \\
J1731-347 & \checkmark & [2.4,27.0] & [2.4,6.1] & 30 & 4.71e-19 & -- \\
G296.1-00.5 & \checkmark & [2.8,28.0] & [3.5,5.1] & 30 & 1.32e-19 & -- \\
MSH 11-61A & \checkmark & [10.0,20.0] & [3.5,11.0] & 16 & -- & -- \\
CTB 37A & \checkmark & [6.0,30.0] & [6.3,12.5] & 15 & -- & -- \\
3C400.2 & \checkmark & [15.0,50.7] & [6.7,7.8] & 30 & -- & -- \\
W63 & -- & [14.1,20.9] & [1.3,3.2] & 78 & -- & -- \\
G065.3+05.7 & -- & $\leq$20.0 & [0.8,1.5] & 270 & -- & -- \\
G156.2+05.7 & -- & [7.0,36.6] & [0.68,3.0] & 110 & -- & --

\end{tabular}

\caption{All SNRs from the set that will not be detectable. The second column indicates whether they will be physically observable from the SWGO site. Fluxes below 1e-19\,TeV\,cm$^{-2}$\,s$^{-1}$ are omitted.}
\label{tab:SNRlist}
\end{table}

\section{Details of MCMC Fitting}
\label{sec:mcmc}
\begin{figure*}
    \centering
    \includegraphics[width=0.999\linewidth]{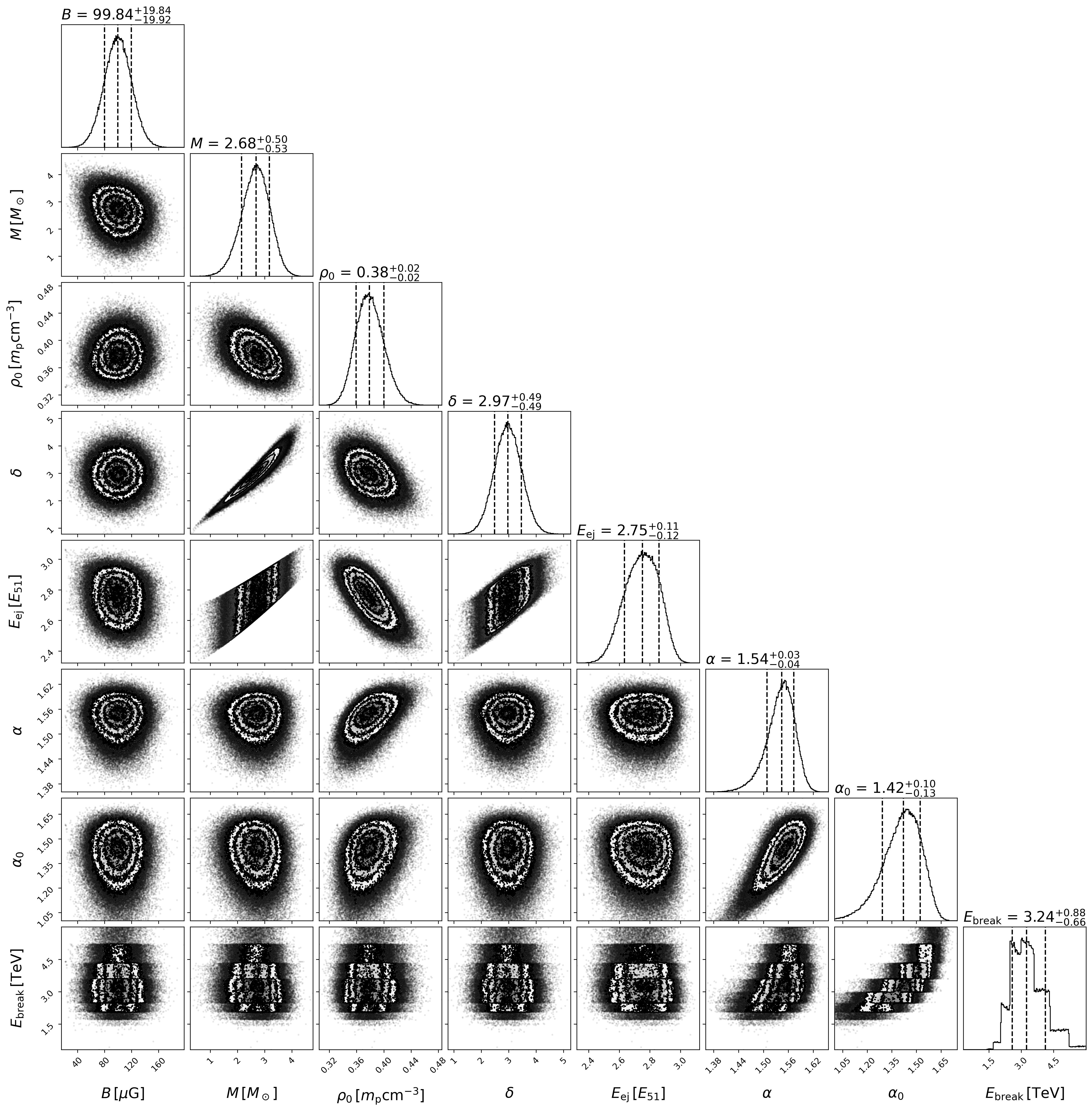}
    \caption{MCMC results for RX J1713.7-3946.}
    \label{fig:RX J1713.7-3946 corner}
\end{figure*}

\begin{figure*}
    \centering
    \includegraphics[width=0.999\linewidth]{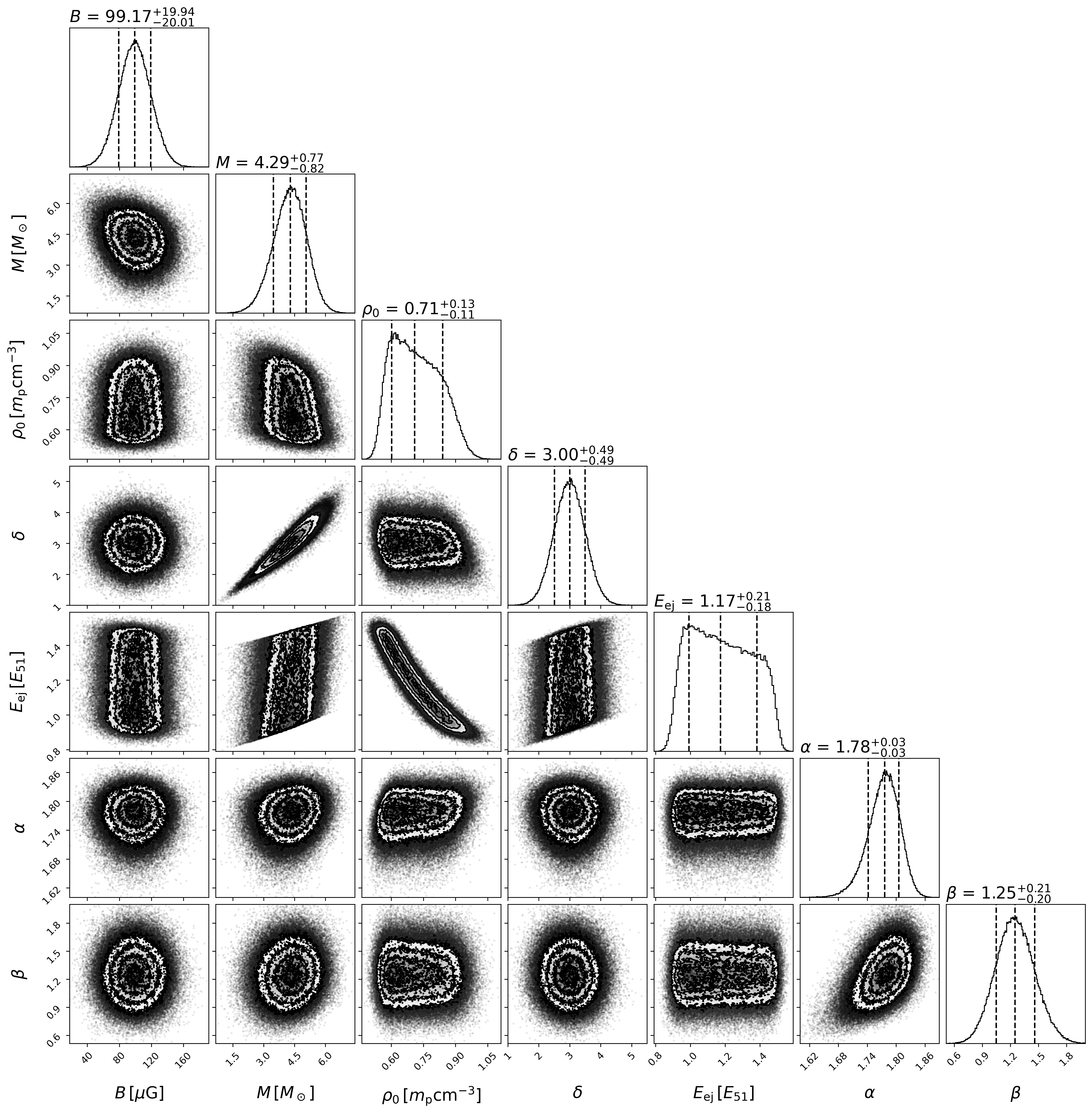}
    \caption{MCMC results for Vela Jr.}
    \label{fig:Vela Jr corner}
\end{figure*}

\begin{figure*}
    \centering
    \includegraphics[width=0.999\linewidth]{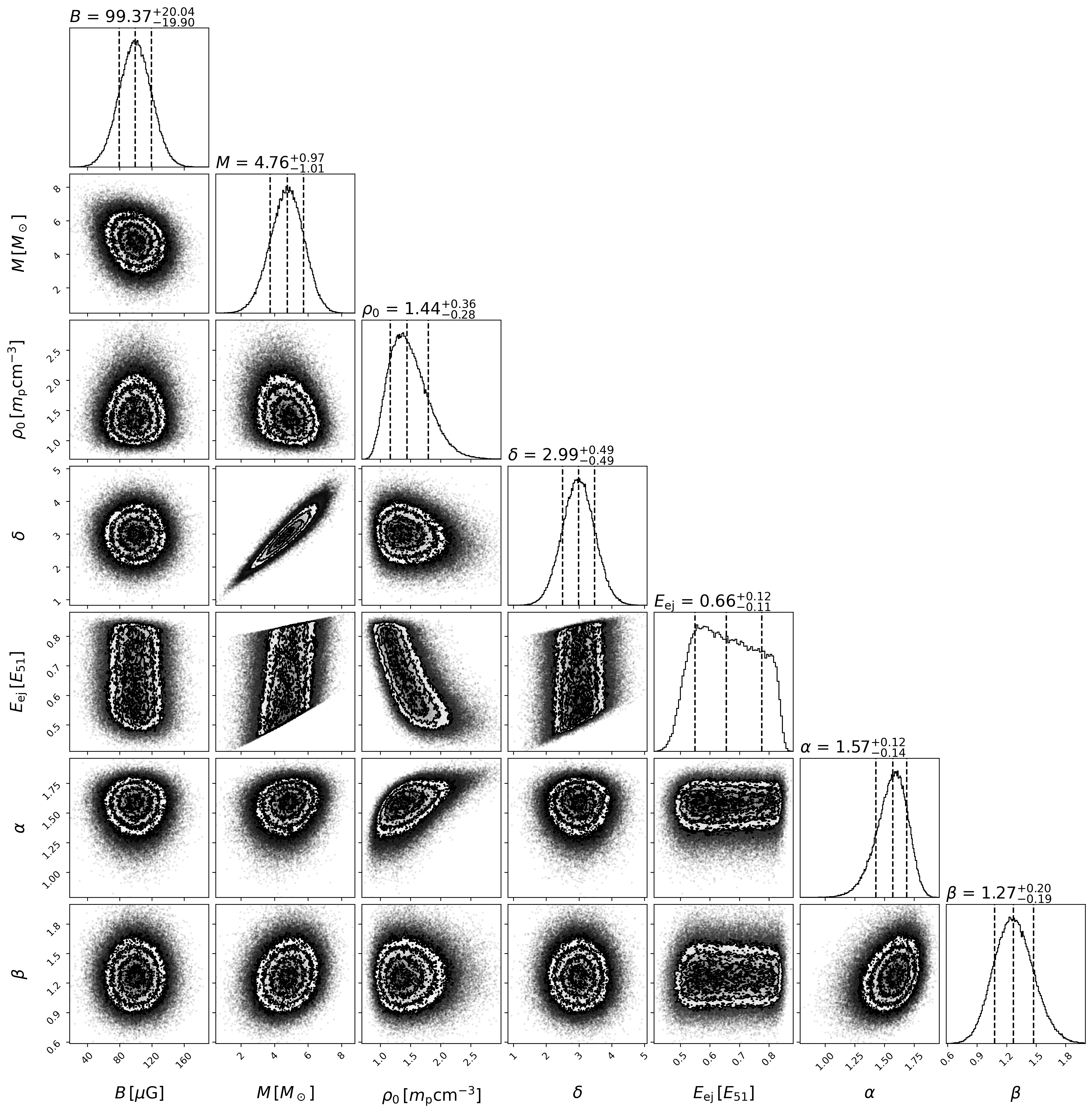}
    \caption{MCMC results for RCW 86.}
    \label{fig:RCW 86 corner}
\end{figure*}

\begin{figure*}
    \centering
    \includegraphics[width=0.999\linewidth]{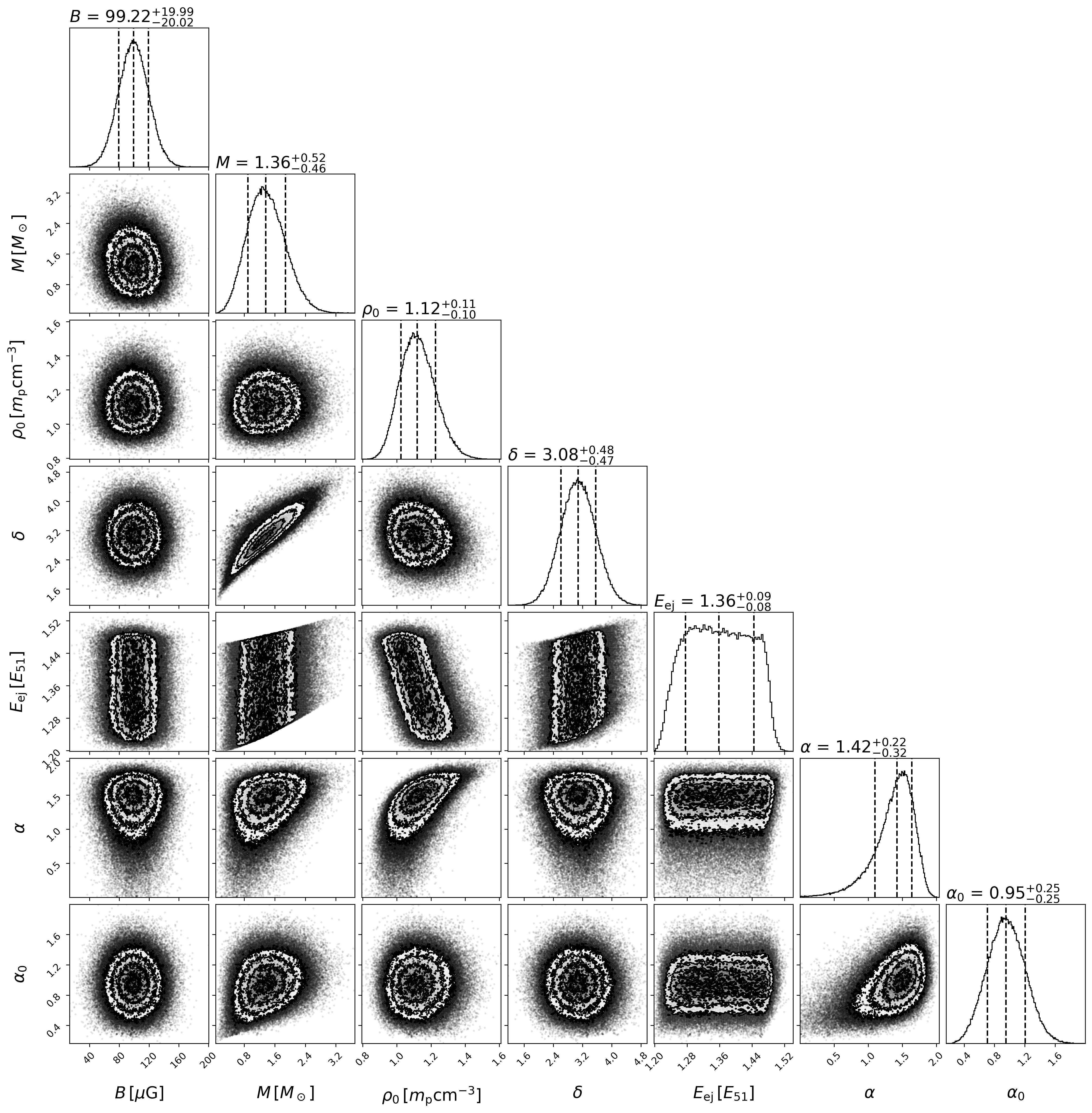}
    \caption{MCMC results for Puppis A.}
    \label{fig:Puppis A corner}
\end{figure*}

\begin{figure*}
    \centering
    \includegraphics[width=0.999\linewidth]{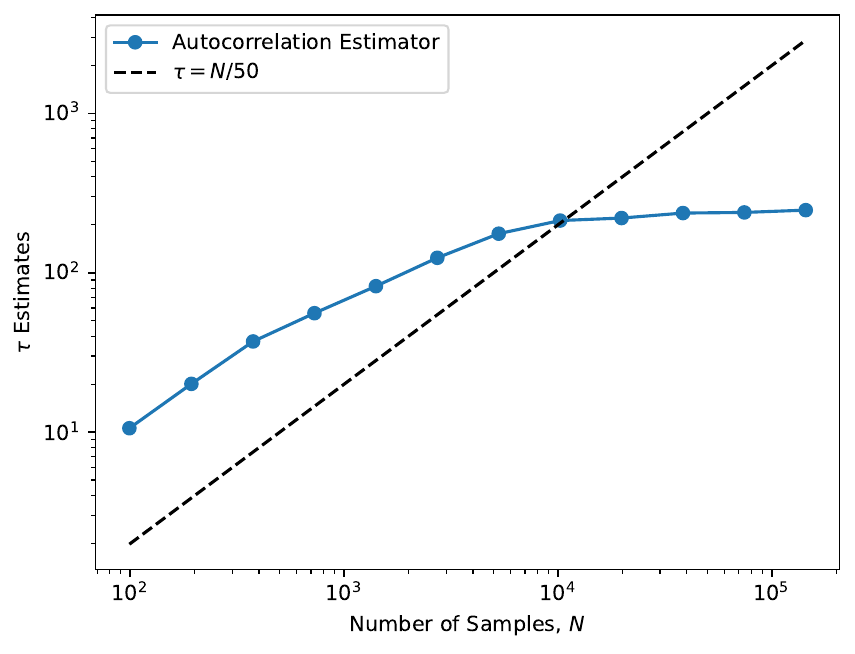}
    \caption{MCMC convergence for RX J1713.7-3946.}
    \label{fig:RX J1713.7-3946 convergence}
\end{figure*}

\begin{figure*}
    \centering
    \includegraphics[width=0.999\linewidth]{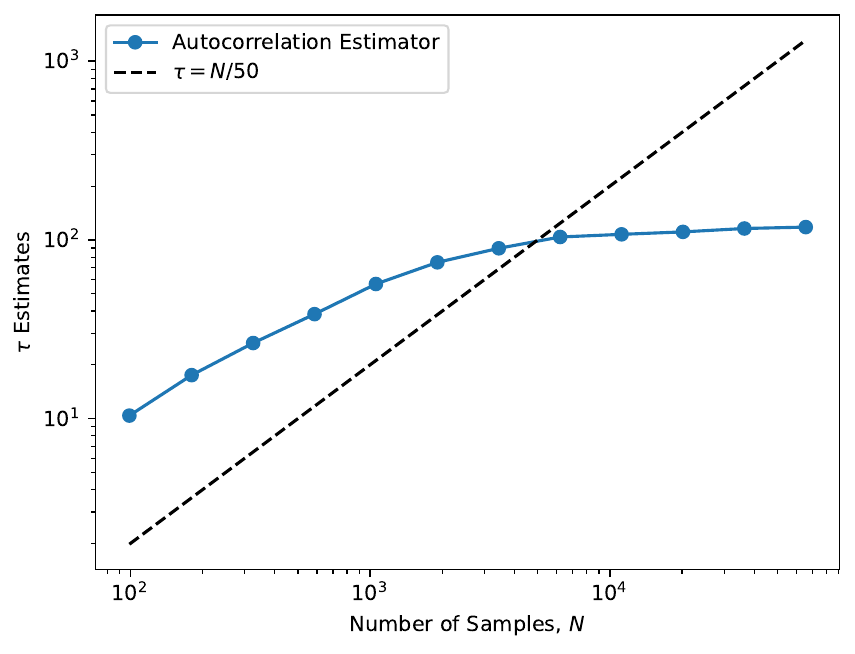}
    \caption{MCMC convergence for Vela Jr.}
    \label{fig:Vela Jr convergence}
\end{figure*}

\begin{figure*}
    \centering
    \includegraphics[width=0.999\linewidth]{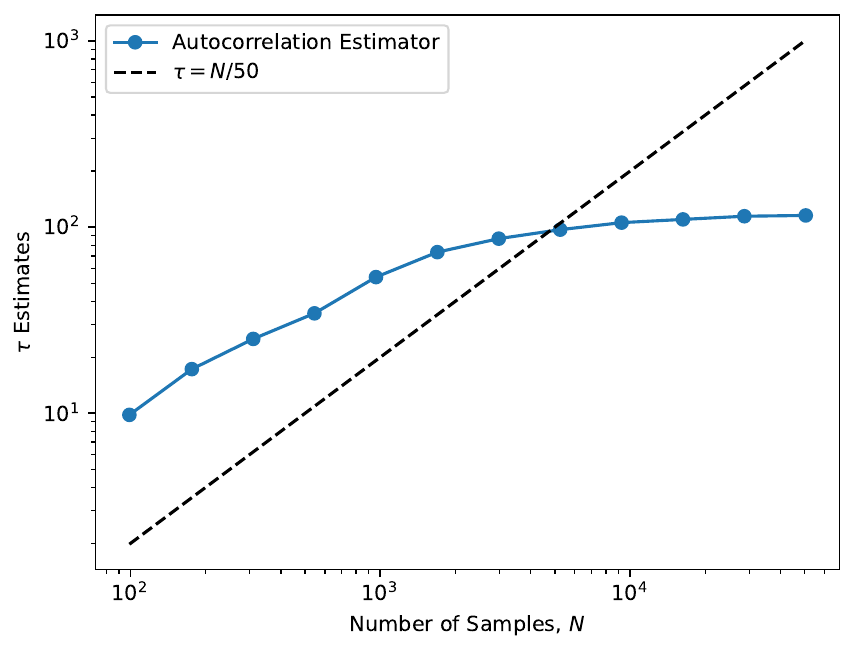}
    \caption{MCMC convergence for RCW 86.}
    \label{fig:RCW 86 convergence}
\end{figure*}

\begin{figure*}
    \centering
    \includegraphics[width=0.999\linewidth]{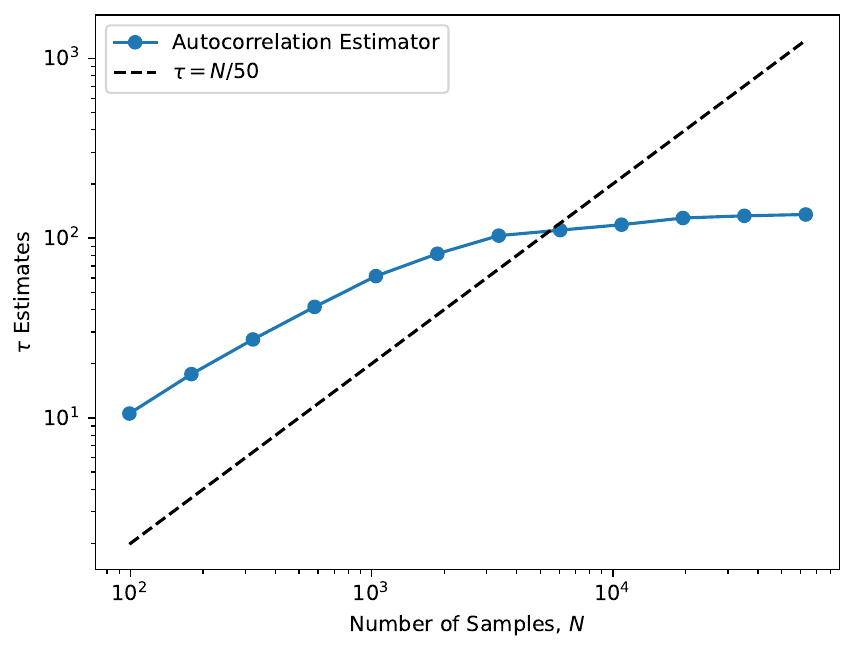}
    \caption{MCMC convergence for Puppis A.}
    \label{fig:Puppis A convergence}
\end{figure*}

MCMC algorithms are a class of techniques to sample probability distributions, and are therefore highly useful for fitting multi-parameter functions to data. Through sampling the posterior probability distributions of these variables around their determined optimum value, they allow one to provide robust uncertainties and to explore covariances between parameters \citep{Hogg_2018}. To perform the MCMC fitting, we follow the recommendation of \citep{Hogg_2018} and use the \textit{scipy} package \citep{2020SciPy-NMeth} to find the best starting position in the multi-dimensional parameter space for the MCMC sampler, for which we use the \textit{emcee} package \citep{Foreman_Mackey_2013} to obtain posterior distributions. These are obtained by having a number of `walkers' performing random steps around the parameter space in order to find the best possible fit to the data. In order to help the MCMC sampler find this optimal fit, it is useful to specify a prior that describes the likely range and distribution of the fitted variables. \added{We detail the priors we use in this work in Table \ref{tab:priors}. }In this appendix, we \added{also }display corner plots for the three sources we consider (Figures \ref{fig:RX J1713.7-3946 corner}, \ref{fig:Vela Jr corner}, \ref{fig:RCW 86 corner} and \ref{fig:Puppis A corner})\replaced{;}{,} these show the 1 and 2-dimensional projections of the posterior parameter distributions, and can be used to visualise covariances between parameters. We also show convergence plots based on the autocorrelation metric described in detail in \citep{Hogg_2018} (Figures \ref{fig:RX J1713.7-3946 convergence}, \ref{fig:Vela Jr convergence}, \ref{fig:RCW 86 convergence} and \ref{fig:Puppis A convergence}) to investigate whether the walkers have had sufficient time to robustly explore the parameter space. All four fits show good evidence for convergence.


\begin{table}[]
\centering
\begin{tabular}{ccccc}
parameter & RX J1713.7-3946 & Vela Jr & RCW 86 & Puppis A \\ \hline
$B$ ($\mu$G) & $\mu=100$, $\sigma=20$ & $\mu=100$, $\sigma=20$ & $\mu=100$, $\sigma=20$ & $\mu=100$, $\sigma=20$ \\[.5mm]
$M_{\rm ej}$ ($M_\odot$) & RP: $a=2$ & RP: $a=1$ & RP: $a=3$ & RP: $a=1.25$ \\[.5mm]
$\rho$ ($m_{\rm p}$cm$^{-3}$) & [0, 10] & [0, 6] & [0, 3] & [0, 5] \\[.5mm]
$\delta$ & $\mu=3$, $\sigma=0.5$ & $\mu=3$, $\sigma=0.5$ & $\mu=3$, $\sigma=0.5$ & $\mu=3$, $\sigma=0.5$ \\[.5mm]
$E_{\rm ej}$ ($E_{51}$) & RP & RP & RP & RP \\[.5mm]
$\alpha_\mathrm{p}$ & -- & [1.6, 2.2] & $\mu=2$, $\sigma=0.25$ & [0, 5] \\[.5mm]
$\beta_\mathrm{p}$ & -- & $\mu=1$, $\sigma=0.25$ & $\mu=1$, $\sigma=0.25$ & $\mu=1$, $\sigma=0.25$ \\[.5mm]
$\alpha$ & [1.35, 4.00] & -- & -- & -- \\[.5mm]
$\alpha_0$ & [1, 5] & -- & -- & -- \\[.5mm]
$E_{\rm break}$ (TeV) & [0, 6] & -- & -- & --
\end{tabular}
\caption{The priors used for each parameter. The intervals stand for uniform priors, $\mu/\sigma$ for gaussian priors. RP stands for ``radius prior'' and is described separately.}
\label{tab:priors}
\end{table}

\paragraph{Radius Prior}
Since the shock radius is described via Equation \ref{eq:radius_sedov_phase}\added{,} which depends on $M_{\rm ej}$ and $E_{\rm ej}$, we can define a prior that constrains the \replaced{modelled}{theoretical} radius to be within the range derived from the given angular size and distance\replaced{, i.e. we check whether the radius for given $M_{\rm ej}$ and $E_{\rm ej}$ estimated at the youngest age is larger than the lower limit of the radius and vice versa at the upper limit.}{.} However, \added{since for some SNRs the range of possible ages is quite large, the possible values for $M_{\rm ej}$ and $E_{\rm ej}$ are quite restricted. So} to let the sampler explore the parameter space \deleted{and not be to restrictive in the first place,} we took a\added{s a uniform prior}\deleted{ multiple of} the
radius range $\overline{R}\pm\Delta R*a$, where $a$ is the factor by which we increase the allowed radius.

\end{document}